\documentclass{aa}

\usepackage{graphicx}
\usepackage{txfonts}
\usepackage[dvipsnames]{xcolor}
\usepackage{xfrac}
\usepackage{chemformula}
\usepackage[colorlinks=true,urlcolor=blue,citecolor=blue,linkcolor=blue]{hyperref}
\usepackage{gensymb}
\usepackage{textgreek}
\usepackage{siunitx}

\begin{document}

   \title{Detection of faculae in the transit and transmission spectrum of WASP-69b}


   \author{D. J. M. Petit dit de la Roche
          \inst{1}
        \and
          H. Chakraborty \inst{1}
          \and
          M. Lendl\inst{1}
          \and
          D. Kitzmann \inst{2,3}
          \and
          A. G. M. Pietrow \inst{4,5}
          \and 
          B. Akinsanmi \inst{1}
          \and
          H. M. J. Boffin \inst{6}
          \and
          Patricio E. Cubillos \inst{7,8}
          \and
          A. Deline\inst{1}
          \and
          D. Ehrenreich\inst{1}
          \and
          L. Fossati\inst{7}
          \and 
          E. Sedaghati \inst{9}
          }

   \institute{Observatoire astronomique de l'Université de Genève, Chemin Pegasi 51, 1290 Versoix, Switzerland
            \and
            Weltraumforschung und Planetologie, Physikalisches Institut, Universität Bern, Gesellschaftsstrasse 6, 3012 Bern, Switzerland 
            \and
            Center for Space and Habitability, Universität Bern, Gesellschaftsstrasse 6, 3012 Bern, Switzerland 
            \and
            Leibniz-Institut für Astrophysik Potsdam (AIP), An der Sternwarte 16, 14482 Potsdam, Germany
            \and
   Centre for mathematical Plasma Astrophysics, Department of Mathematics, KU Leuven, Celestijnenlaan 200B, B-3001 Leuven, Belgium
   \and
   European Southern Observatory (ESO), D-85748 Garching, Germany
   \and
   Space Research Institute, Austrian Academy of Sciences,
    Schmiedlstrasse 6, A-8042, Graz, Austria 
    \and
    INAF -- Osservatorio Astrofisico di Torino,
    Via Osservatorio 20, 10025 Pino Torinese, Italy 
   \and
   European Southern Observatory (ESO), Av. Alonso de C\'{o}rdova 3107, 763 0355 Vitacura, Santiago, Chile
   \\
              \email{dominique.petit@unige.ch}
             }

   \date{Received XXX; accepted XXX}

 
  \abstract
   {Transmission spectroscopy is a powerful tool for understanding exoplanet atmospheres. At optical wavelengths, it makes it possible to infer the composition and the presence of aerosols in the atmosphere. However, unocculted stellar activity can result in contamination of atmospheric transmission spectra by introducing spurious slopes and molecular signals. }
   {We aim to characterise the atmosphere of the transiting exoplanet WASP-69b, a hot Jupiter orbiting an active K star, and characterise the host star's activity levels. }
   {We obtained three nights of spectrophotometric data with the FORS2 instrument on the VLT, covering a wavelength range of 340-1100\,nm. These were divided into 10\,nm binned spectroscopic light curves, which were fit with a combination of gaussian processes and parametric models to obtain a transmission spectrum. We performed retrievals on the full spectrum with combined stellar activity and planet atmosphere models. }
   {We directly detect a facula in the form of a hot spot crossing event in one of the transits and indirectly detect unocculted faculae through an apparently decreasing radius towards the blue end of the transmission spectrum. We determine a facula temperature of $\Delta T=+644^{+427}_{-263}$\,K for the former and a stellar coverage fraction of around 30\% with a temperature of $\Delta T=+231\pm72$\,K for the latter. The planetary atmosphere is best fit with a high-altitude cloud deck at 1.4\,mbar that mutes atomic and molecular features. We find indications of water and ammonia with log(H$_2$O)=$-2.01^{+0.54}_{-0.86}$ and log(NH$_3$)=$-3.4^{+0.96}_{-5.20}$ respectively and place 3\textsigma\ upper limits on TiO ($<10^{-7.65}$) and K ($<10^{-7}$). We see no evidence of Na which we attribute to the presence of clouds. }
   {
   The simultaneous multi-wavelength observations allow us to break the size-contrast degeneracy for facula-crossings, meaning we can obtain temperatures for both the directly and indirectly detected faculae, which are consistent with each other.}

   \keywords{ Methods: observational, Planets and satellites: individual: WASP-69b, Stars: activity
               }

   \maketitle
%

\section{Introduction}

Transmission spectroscopy is a powerful and effective way of probing the atmospheres of exoplanets. The wavelength dependence of the planet radius reveals features of not only atoms and molecules \citep[e.g. sodium and water;][]{Charbonneau2002,Deming2013,Nikolov2016,Nikolov2018,Sedaghati2017,Carter2020}, but also of clouds and hazes through broad trends in the optical and near-infrared spectra \citep[e.g.][]{Pont2008,Lecavelier2008,Sing2016,Gao2020,Spyratos2021}. 

However, interpretation of these spectra is complicated by the presence of stellar inhomogeneities that affect the observed features \citep[e.g.][]{Oshagh2013,McCullough2014}. These inhomogeneities can take the form of cooler and darker areas called sunspots on the Sun, or starspots on other stars \citep{Solanki2003}. Hotter and brighter areas are called \emph{plages} if they are brighter in the chromosphere 
and \emph{network} if they are brightest in the photosphere. Plages and network are both also referred to as \emph{faculae}, 
although there is some discussion on the exact definition of this term \citep[e.g.][]{Buehler2019, Chintzoglou2021, Cretignier2024}. In this work we use faculae as a catch-all for bright areas for convenience. 

When active areas occur in the path of the planet, they appear as bumps or dips in the light curves for spots and faculae respectively, which can affect the transit fit and therefore the retrieved planet radius and transit midpoint, even to the point of mimicking transit timing variation signals \citep{Czesla2009,Oshagh2014,Ioannidis2016}. 
While there have been numerous examples of spot crossings in the literature \citep[e.g.][]{Silva2003,Nutzman2011,LibbyRoberts2023}, there appear to be a limited number of facula crossings, which have also been observed more recently \citep{Mohler2013,Kirk2016,Zaleski2019,Zaleski2020,Jiang2021,Baluev2021}.

This is surprising, as faculae are not expected to be rare. In particular there may be a transition from spot-dominated to facula dominated stellar variability at 15-25 day rotation periods \citep{Montet2017}. Combined with rotation rates from Kepler stars, this would mean that roughly half of all stars should be facula dominated \citep{Reinhold2023}. Faculae should therefore be especially common in later type stars that are generally considered to be more active and have longer rotation periods \citep{Nielsen2013}. 

While activity in the planets path can be accounted for in the light curve fitting process \citep[for example with codes like \texttt{PyTranSpot} and SOAP-T or with gaussian processes;][]{Oshagh2013,Juvan2018,Bruno2018}, activity in unocculted areas of the star is harder to correct for. Unocculted activity features cause what is known as the transit light source effect \citep[TLS effect;][]{Rackham2018,Pont2008,Pont2013,McCullough2014}. In this case, the area of the stellar surface traversed by the planet is no longer representative of the full stellar disk, as is otherwise assumed. Colder and dimmer spots outside the transit chord mean that the luminosity outside the chord is lower than inside it and the planet blocks a larger fraction of star light than assumed, leading to a larger observed radius. On the other hand, hotter and brighter faculae have the opposite effect, resulting in a smaller apparent radius. The TLS effect is strongest in the optical where the emission differences as a result of temperature difference are greatest, resulting in an added slope in the transmission spectra. This complicates the interpretation of the transmission spectra, since the TLS effect can mimic or hide slopes induced by aerosols \citep[e.g.][]{Murgas2020}. Additionally, unocculted spots can induce features mimicking atmospheric constituents such as sodium, water and TiO in the transmission spectrum, making such features hard to interpret without sufficient wavelength coverage \citep[e.g.][]{Mallia1970,Murgas2020,Moran2023}.

At high resolution, the equivalent width of lines that are sensitive to stellar activity can be used to estimate the active region coverage fraction \citep{Guilluy2020,Dineva2022}, but at low resolution retrievals of the planetary transmission spectrum are necessary to quantify the stellar activity. This can be done either through a combined stellar activity and planetary atmosphere retrieval, or by estimating how well the transmission spectrum can be explained by stellar activity alone under the assumption that the planet has no atmosphere. 

In this paper, we present the optical transmission spectrum of WASP-69 obtained with the FORS2 instrument on the VLT \citep{Appenzeller1998}. To account for the contamination of the transmission spectrum by the TLS effect, we include it in the Bern Atmospheric Retrieval code (\textsc{BeAR})\footnote{\textsc{BeAR} is available in the GitHub repository: \url{https://github.com/newstrangeworlds/bear}} through parameters for the photospheric temperature and spot and faculae coverage fractions and temperatures.  


WASP-69b is an inflated Saturn-mass planet ($0.26\,\mathrm{M_{Jup}}, 1.06\,\mathrm{R_{Jup}}$) on a 3.87\,day orbit around a $\sim$1\,Gyr old active K-type main sequence host star \citep[][]{Anderson2014}. It is well characterised in the optical and near-infrared. Eclipse observations with Spitzer indicate a 100x solar metallicity atmosphere with a brightness temperature of $971\pm20$\,K \citep{Wallack2019}. High resolution studies have detected sodium in the atmosphere, although not consistently and stronger in the D2 line than the D1 line \citep{Casasayas2017,Deibert2019,Khalafinejad2021}. They have also identified a comet-like tail of escaping helium formed from atmospheric erosion due to the proximity to the star \citep{Nortmann2018,Tyler2024,Guilluy2024}. This tail appears to be highly variable, with at least one study not detecting it at all \citep{Vissapragada2020}, possibly as a result of stellar activity \citep{Guilluy2024}. \citet{Guilluy2022} also detect CH$_4$, NH$_3$, CO, C$_2$H$_2$, and H$_2$O. Spectrophotometric transit observations have also detected the presence of water in the atmosphere and detected aerosols in the form of clouds and hazes \citep{Fisher2018,Murgas2020,Estrela2021,Ouyang2023}. Stellar activity has complicated these observations, with \citet{Murgas2020}, \citet{Estrela2021} and \citet{Ouyang2023} all showing that the slopes they detect in the transmission spectrum can be caused either by aerosols or by stellar activity. Most recently, JWST eclipse observations at 2-11\,\textmu m have confirmed the presence of CO, CO$_2$ and H$_2$O \citep{Schlawin2024}.

We present our observations in Section \ref{sec:obs} and describe the data analysis and simultaneous stellar activity and atmospheric retrievals in Section \ref{sec:methods}. In Section \ref{sec:results} we show the data reduction and retrieval results. The discussion and conclusions are presented in Section \ref{sec:disc} and Section \ref{sec:conc} respectively.

\section{Observations}
\label{sec:obs}

\begin{table*}
\caption{Observation details}             
\label{tab:observations}      
\centering                          
\begin{tabular}{c c l c c c c c}        
\hline\hline                 
Instrument & date & grism & filter & wavelengths (nm) & exposure time (s) & airmass & average seeing (") \\    
\hline                        
   
    FORS2  & 18-07-2017 & 600RI & GG435 & 510 - \phantom{0}850 & 228 x 15 & 1.795 - 1.061 & 1.8\\  
            & 14-08-2017 & 600B  & -    & 340 - \phantom{0}610 & 184 x 40 & 1.061 - 2.357 & 0.7\\ 
            & 18-08-2017 & 600z  & OG590 & 740 - 1100 & 152 x 20 & 1.436 - 1.061 & 0.7\\ \hline 
    ECAM & 18-07-2017 &   \phantom{aaa}-   &       Johnson-$B$&           352 - 518&           188 x [50-120]&               2.165 - 1.137&  1.2  \\
    & 14-08-2017  &   \phantom{aaa}-   &       Johnson-$B$&           352 - 518&           160 x 110 \phantom{000|}&               1.146 - 2.546&    1.1\\
    
    & 18-08-2017 &   \phantom{aaa}-   &       Johnson-$B$&           352 - 518&           165 x [70-100]&               1.871 - 1.130&    0.9\\

\hline                                   
\end{tabular}
\end{table*}

We observed three transits of WASP-69b on different nights in July and August of 2017. We obtained simultaneous photometric and spectroscopic measurements with different facilities. Details for each night are displayed in Table \ref{tab:observations}. 

\subsection{FORS2 low-resolution spectroscopy}
Spectroscopic observations were done with the FOcal Reducer/low dispersion Spectrograph 2 \citep[FORS2;][]{Appenzeller1998} instrument on the Very Large Telescope (VLT) as part of the CHEWIE programme (Clouds, Hazes and Elements vieWed on gIant Exoplanets; 099.C-0189). The instrument was used in the spectroscopic mask (MXU) mode, allowing for the simultaneous observation of WASP-69 and the reference star TYC 5187-1718-1 through the use of a custom, laser-cut mask that is inserted into the focal plane. The bulk of the observations were taken with straight slits with widths of 13\arcsec to 18\arcsec and lengths of 35\arcsec. A number of calibration spectra were taken with narrower slit widths of 0.4\arcsec at the beginnings and ends of the observations. Three different grisms and corresponding order-separation filters were used to obtain a continuous wavelength coverage from 340\,nm to 1100\,nm, with a single transit being observed in each grism/filter combination. The nominal resolving powers are R$\sim$780, 1000, and 1260 for the B, RI, and z grisms respectively, although due to the use of larger slit widths, the effective resolutions of the observations may be lower. 

While the seeing was good on the last two nights, it was significantly worse on the first night, possibly causing the larger amplitude of the correlated noise in those observations, as can be seen in Fig. \ref{fig:wl_curves}. The scatter at the beginning of the night is caused mostly by a high airmass, as is the scatter at the end of the second night. 

\subsection{Simulteanous EulerCam Photometry}

In parallel with our VLT/FORS2 observations, we obtained broadband photometry with EulerCam \citep{Lendl2012}. EulerCam or ECAM is a 4k $\times$ 4k CCD camera installed at the 1.2 meter Euler telescope at La Silla, Chile. The observations were taken with the Johnson-\textit{B} filter (352-518 nm) to maximise the signal arising from any spot or faculae crossings. In addition, taking simultaneous observations from la Silla and Paranal helps us to effectively disentangle weather and instrument systematic effects from astrophysical signals.


\section{Methods}
\label{sec:methods}

\subsection{Spectroscopy}
We use a custom Python pipeline we call HANSOLO (atmospHeric trANsmission SpectrOscopy anaLysis cOde) that was especially developed for the CHEWIE dataset to perform the data reduction and analysis of the FORS2 data. We analyse each night separately to obtain individual transmission spectra. Offsets between the transmission spectra of the different nights are corrected so that the B and z grism spectra align with the RI grism observations in the overlapping regions. The RI grism observations were chosen as a reference as it is the only mode with wavelength overlap with both the other modes. The spectra are then combined for a simultaneous stellar activity and atmospheric retrieval. 

\subsubsection{Spectral extraction and background removal}
We follow the data reduction set out in \citet{Lendl2016,Lendl2017} for FORS2 data. Standard bias and flatfield corrections are applied and the two-dimensional detector images are wavelength calibrated using spectra taken with the ThAr lamps (performed with an identical mask to the science mask, but with \SI{1}{\arcsec} slit widths). Cosmic rays are identified during this step using the LACOSMIC algorithm \citep{vanDokkum2001}. The spectral extraction is done on a column-by-column basis, with the middle of the trace being identified in each column by fitting a Moffat function. The background is removed by fitting a linear trend to twenty pixels at each edge of the slit and removing that from the column. The flux is then obtained by integrating over eleven apertures around the center of the trace with widths of 25-75 pixels. We remove any outliers beyond 5 to 10\,\textsigma, depending on the quality of the data, from the extracted spectra. These values were chosen to remove outliers from cosmic rays or broken pixels, without removing sections of the spectra with high variation from seeing or other observational causes. Overall, around 0.1\% of the final data points are rejected. We align all spectra with the first observed spectrum by cross-correlating 3-5 lines at different wavelengths depending on availability of lines in the grism and fitting a second order polynomial to the relative shifts. The spectra are then interpolated to fit the same wavelength grid. This accounts for both shifts and stretching of the trace on the detector due to varying observing conditions during the night. We then apply a second wavelength correction based on a PHOENIX model stellar spectrum with the same temperature and log(g) as WASP-69. 

\begin{figure}
    \centering
    \includegraphics[width=0.5\textwidth,trim=1cm 0cm 1.5cm 0cm]{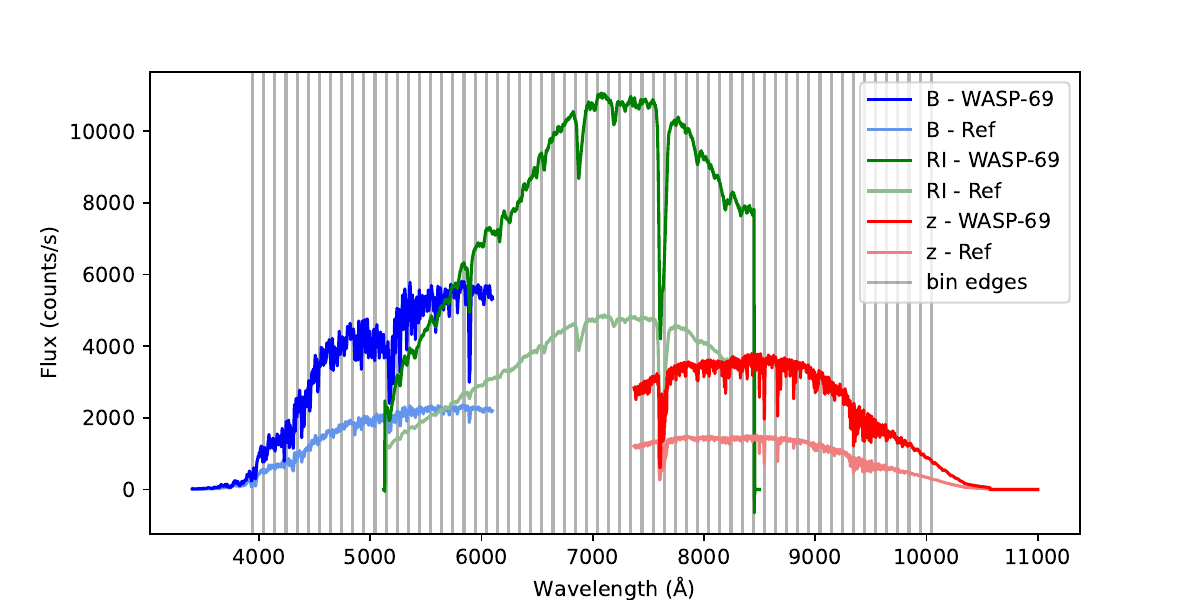}
    \caption{Observed spectra of WASP-69 and reference star TYC 5187-1718-1 in the B (blue), RI (green) and z (red) grisms. Gray vertical lines indicate the edges of the spectroscopic light curve bins. }
    \label{fig:enter-label}
\end{figure}

\subsubsection{White light curves}
"White", i.e. wavelength-integrated, light curves are obtained for the different apertures by integrating over the full spectra. The aperture with the lowest median absolute deviation (MAD) is selected for the rest of the analysis. These correspond to apertures of 40, 70, and 65 pixels for the B, RI, and z grisms respectively. 

To obtain the system parameters we first fit the white light curves with a transit model and a gaussian process \citep[GP;][]{Rasmussen2006,Gibson2012,Gibson2014} with a Matern 3/2 kernel, using the CONAN transit fitting code \citep{Lendl2017,Lendl2020}. We fix the period and the eccentricity to the literature values \citep[3.8681382$\pm$0.00000017\,days and 0.00$\pm$0.05 respectively;][]{Anderson2014} as we are analysing the different nights separately and so only have one transit per fit. We also fix the limb darkening parameters to values obtained with the \texttt{LDCU} code as they are highly degenerate with the impact parameter. \texttt{LDCU}\footnote{\url{https://github.com/delinea/LDCU}} is a modified version of the python routine implemented by \citet{Espinoza2015} that computes the limb-darkening coefficients and their corresponding uncertainties using a set of stellar intensity profiles accounting for the uncertainties on the stellar parameters. In this case the stellar parameters were obtained from \citet{Anderson2014}: $T_\mathrm{eff}=4715\pm50$, $log(g)=4.5\pm0.15$ and $[Fe/H]=0.114\pm0.077$. The stellar intensity profiles are generated based on two libraries of synthetic stellar spectra: ATLAS \citep{Kurucz1979} and PHOENIX \citep{Husser2013}. We adopted the Kipping parametrisation of the quadratic limb darkening law for computational efficiency \citep{Kipping2013}. The planet radius, the impact parameter and the transit midpoint and duration are left as free parameters, as well as the GP hyperparameters. CONAN fits also include a flux offset and a jitter parameter that estimates the extra white noise in the data. The full set of priors and the results for each night can be found in Table \ref{tab:wl_fits}. 

\begin{figure*}
      \includegraphics[width=\textwidth]{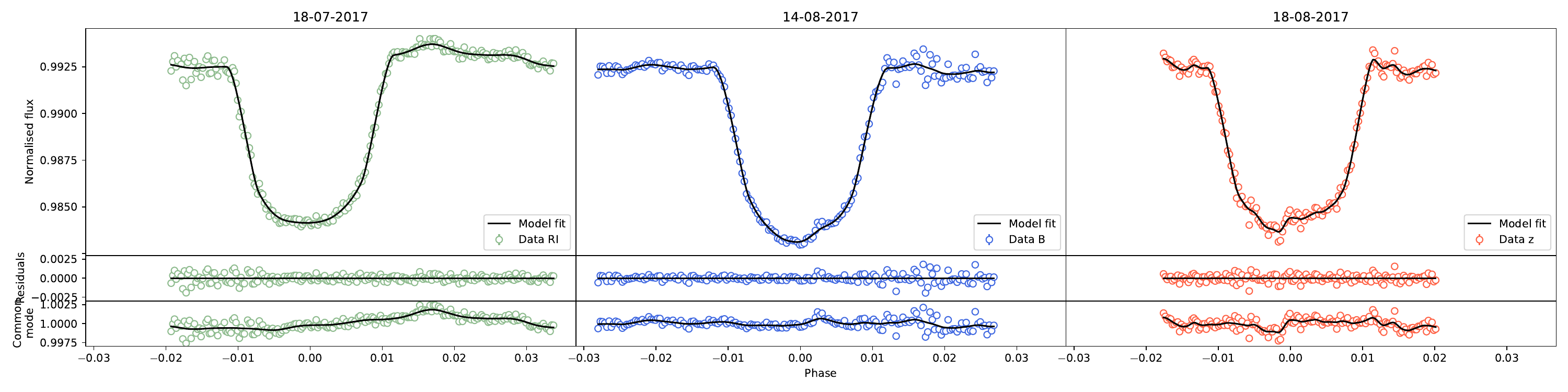} 
      \caption{The top row shows the white light curves and fits for all three nights: the first night observations were done in the RI grism (green, left), the second in the B grism (blue, middle) and the third in the z grism (red, red). Data points are indicated with open circles while the solid lines indicate the fitted transit and GP model. The second row shows the residuals for each of the light curves and the third row shows the obtained common mode noise model in colour with the GP fit overplotted in black. The GP fits and common noise models for the B and z grisms clearly show the crossing of the respective active regions.} 
      
         \label{fig:wl_curves}
   \end{figure*}

Figure \ref{fig:wl_curves} shows the white light curves in the upper panels, with the best fits overplotted in black. The middle panels show the residuals and the bottom panels show the residual after just the transit models have been removed, with the GP contributions to the models overplotted in black. Since two of the curves contain signs of active region crossing, we also fit the B and z grism white light curves with the data points around the crossings of active regions excluded, but find that the obtained parameters are not significantly different. 

\begin{table*}
    \caption{White light fits for the three different nights. The same priors were used for each white light fit, with the exception of the transit midpoint, which was adjusted for the observation time. The GP parameters for the lengthscale (log $\rho$) and amplitude (log $\sigma$) of the correlated noise and the jitter parameter (also in log-scale) have uniform priors. In the transit midpoint prior $t$ is used to indicate the date-dependant estimated (by eye) value used as prior. The second part of the table contains the physical properties of the system as derived from the fitted parameters.}
    \label{tab:wl_fits}
    \centering
    \begin{tabular}{llccc}
    \hline \hline
    Parameter & Prior & \phantom{000}B & \phantom{000}RI & \phantom{000}z \\
    \hline
    Planet radius, R$_\mathrm{p}$ $\mathrm{[R_*]}$ & $\mathcal{U}(0,1)$ & \phantom{000}0.131\phantom{0} $\pm$ 0.002\phantom{0} & \phantom{000}0.126\phantom{0} $\pm$ 0.003\phantom{0} & \phantom{000}0.125\phantom{0} $\pm$ 0.002\phantom{0} \\
    Impact parameter, b  & $\mathcal{U}(0,1)$ & \phantom{000}0.68\phantom{00} $\pm$ 0.02\phantom{00} & \phantom{000}0.66\phantom{00} $\pm$ 0.02\phantom{00} & \phantom{000}0.66\phantom{00} $\pm$ 0.03\phantom{00} \\
    Transit midpoint, T$_\mathrm{0}$ [days] & $\mathcal{N}(t, 0.05)$ & 7980.7505 $\pm$ 0.0003 & 7953.6739 $\pm$ 0.0003 & 7984.6184 $\pm$ 0.0004 \\
    Transit duration, T$_\mathrm{14}$ [days] & $\mathcal{N}(0.1,0.05)$ & \phantom{000}0.093\phantom{0} $\pm$ 0.001\phantom{0} & \phantom{000}0.089\phantom{0} $\pm$ 0.001\phantom{0} & \phantom{000}0.089\phantom{0} $\pm$ 0.001\phantom{0} \\
    Offset & $\mathcal{U}(-1,1)$ & \phantom{000}0.0058 $\pm$ 0.0002 & \phantom{000}0.0057 $\pm$ 0.0007 & \phantom{000}0.0049 $\pm$ 0.0003 \\
    log $\rho$ & $\mathcal{U}(-20,0)$ & \phantom{00}-4.6\phantom{000} $\pm$ 0.6\phantom{000} & \phantom{00}-3.4\phantom{000} $\pm$ 0.4\phantom{000} & \phantom{00}-5.1\phantom{000} $\pm$ 0.6\phantom{000} \\
    log $\sigma$   & $\mathcal{U}(-10,-2)$ & \phantom{00}-7.8\phantom{000} $\pm$ 0.3\phantom{000} & \phantom{00}-6.9\phantom{000} $\pm$ 0.5\phantom{000} & \phantom{00}-7.4\phantom{000} $\pm$ 0.4\phantom{000} \\
    log Jitter  & $\mathcal{U}(-15,-4)$ & \phantom{00}-7.56\phantom{00} $\pm$ 0.07\phantom{00} & \phantom{00}-7.62\phantom{00} $\pm$ 0.05\phantom{00} & \phantom{00}-7.54\phantom{00} $\pm$ 0.08\phantom{00} \\ 
    
    \hline
    Planet radius, R$_\mathrm{p}$ [R$_\mathrm{Jup}$] & - & \phantom{000} 1.04 \phantom{00} $\pm$ 0.04 \phantom{00} & \phantom{000}0.99\phantom{00} $\pm$ 0.04\phantom{00} & \phantom{000}0.99\phantom{00} $\pm$ 0.04\phantom{00} \\
    Semi-major axis, a [au] & - & \phantom{000}0.045\phantom{0} $\pm$ 0.002\phantom{0} & \phantom{000}0.048\phantom{0} $\pm$ 0.002\phantom{0} & \phantom{000} 0.048\phantom{0} $\pm$ 0.003\phantom{0} \\
    Inclination, i [\degree] & - & \phantom{00}86.8\phantom{000} $\pm$ 0.2\phantom{000} & \phantom{00}87.0\phantom{000} $\pm$ 0.2\phantom{000} & \phantom{00} 87.0\phantom{000} $\pm$ 0.3\phantom{000} \\
    \hline
    \end{tabular}
\end{table*}

\subsubsection{Spectroscopic light curves}
We also produce spectroscopic light curves by binning the spectra along the wavelength axis with a bin size of 10\,nm, resulting in 27 light curves in the B grism, 34 light curves in the RI grism and 33 light curves in the z grism.

We obtain a separate common noise model for each night by dividing the white light curve by the transit component of the fitted model. Because of this, the common noise model contains both correlated and uncorrelated noise contributions. This model is then divided out of the corresponding spectroscopic light curves to remove any noise features common to all light curves. Both the original and detrended light curves are shown in Fig. \ref{fig:spectroscopic-B}, Fig. \ref{fig:spectroscopic-RI} and Fig. \ref{fig:spectroscopic-z}, along with the common noise models for the B, RI and z grism observations respectively. 

The spectroscopic light curves are fit with a combined transit and baseline model to account for any weather and instrumental effects still present in the data. For the transit model the \citet{Mandel2002} model is used. Only the planetary radius is fit, while the transit midpoint and duration and the impact parameter are fixed to the values obtained in the white light fits. Limb darkening parameters are calculated for each bin to account for the change in limb darkening with wavelength, but are kept fixed as they are not well constrained when left free. 

For the baseline model for each night, we tried various combinations of polynomials with time, airmass, spectrum full-width-half-maximum (FWHM), sky background and trace position. More complicated models were only accepted over simpler models if the Bayesian Information Criterion (BIC) indicated a significantly higher probability ($|\Delta_{BIC}|>8$). This resulted in a linear polynomial with the spectrum FWHM for all grisms and a polynomial with time that was second order in the B grism and first order in the RI and z grisms. These were chosen over another GP model to prevent overfitting after the removal of the common noise model. Instead, we estimate the residual white and red noise in the light curves by calculating the white and red noise factors for each light curve \citep[$\beta_w$ and $\beta_r$ respectively;][]{Winn2008,Gillon2010}. $\beta_w$ is simply the root-mean-square (RMS) of the residuals over the mean of the errors, while $\beta_r$ compares the RMS of the binned photometric residuals to the RMS over the entire dataset. The errors of the photometric light curves are then inflated with $CF=\beta_w \times \beta_r$ and the light curves are fit again to obtain the errorbars of the final spectrum. These fits and their residuals are plotted in the second and third columns of Fig. \ref{fig:spectroscopic-B}, Fig. \ref{fig:spectroscopic-RI} and Fig. \ref{fig:spectroscopic-z}.  


\subsection{Atmospheric retrievals}

To characterise the observations we performed atmospheric retrieval calculations using the open-source Bern Atmospheric Retrieval code (\textsc{BeAR})\footnote{\textsc{BeAR} is available in the GitHub repository: \url{https://github.com/newstrangeworlds/bear}}. This code is an updated version of the retrieval code previously named \textsc{Helios-r2} \citep{Kitzmann2020ApJ...890..174K}. \textsc{BeAR} employs the \textsc{MultiNest} \citep{Feroz2008MNRAS.384..449F} library to perform the parameter space exploration using Bayesian nested sampling \citep{Skilling2004AIPC..735..395S}. The number of live sampling points was 4000 for all retrieval calculations performed in this study.
In this work we used the transmission spectroscopy forward model of \textsc{BeAR}. It assumes an isothermal atmosphere with a temperature $T$ and a surface gravity $\log g$. For this study, the atmosphere extended from a pressure of 10 bar at the bottom to $10^{-6}$ bar at the top and is divided into 100 atmospheric layers, distributed equidistantly in logarithmic pressure space. The planet's radius at the bottom of the atmosphere, $R_\mathrm{p}$, was used as a free parameter. 

The abundance profiles of the chemical species were assumed to be constant with pressure, described by a single mixing ratio $x_i$ per considered species. In the retrievals we used the mixing ratios of water (\ch{H2O}), titanium oxide (\ch{TiO}), ammonia (\ch{NH3}), sodium (Na), and potassium (K) as free parameters. While gaseous TiO might not be expected at the relatively cool equilibrium temperature of WASP-69b \citep[$T_\mathrm{eq}=963\pm18$;][]{Anderson2014}, it is nevertheless included in the retrieval due to a potential detection of it by \citet{Ouyang2023}. Other molecules, such as for example methane or hydrogen cyanide, were initially considered as well, but were unconstrained and, therefore, removed from the retrieval calculations.

Opacity tables for \ch{H2O}, \ch{TiO}, and \ch{NH3} were computed by the open-source \textsc{HELIOS-K}\footnote{\textsc{HELIOS-K} is available in the GitHub repository \url{https://github.com/exoclime/HELIOS-K}. The computed opacities can be downloaded from the DACE platform \url{https://dace.unige.ch}.} code \citep{Grimm2015ApJ...808..182G, Grimm2021ApJS..253...30G} based on line-list data published by \citet{Polyansky2018MNRAS.480.2597P}, \citet{McKemmish2019MNRAS.488.2836M}, and \citet{AlDerzi2015JQSRT.161..117A} respectively. The opacities for K and Na are based on the Kurucz line list \citep{Kurucz2017CaJPh..95..825K}, with additional data for the description of the resonance line wings based on the work by \citet{Allard2012A&A...543A.159A}, \citet{Allard2016A&A...589A..21A}, and \citet{Allard2019A&A...628A.120A}.

For this study, we added a stellar contamination module to \textsc{BeAR}, following the description by \citet{Rackham2018}. Thus, we modelled the transmission spectrum under the impact of stellar contamination due to spots and faculae according to
\begin{equation}
    D_{\lambda, \mathrm{obs}} = \frac{D_\lambda}{1 - f_{\mathrm{spot}} \left( 1 - \frac{F_\mathrm{spot}(\lambda)}{F_\mathrm{phot}(\lambda} \right)
    - f_{\mathrm{fac}} \left( 1 - \frac{F_\mathrm{fac}(\lambda)}{F_\mathrm{phot}(\lambda} \right)}
\end{equation}
where $D_\lambda$ and $D_{\lambda,obs}$ are the true and observed transit depths, $f_{\mathrm{fac}}$ and $f_{\mathrm{spot}}$ are the spot and faculae covering fractions, $F_\mathrm{phot}$ is the photospheric spectrum of the star, and $F_\mathrm{spot}$ and $F_\mathrm{fac}$ are the spectra of the spots and faculae, respectively.
We sample the corresponding spectra from the spectral library by \citet{Husser2013} based on the PHOENIX stellar atmosphere model. For a given set of stellar parameters, surface gravity $\log g_*$, stellar effective temperature $T_{\mathrm{eff},*}$, and metallicity [Fe/H], we perform a three-dimensional interpolation within the PHOENIX grid to obtain the stellar photospheric spectrum $F_\mathrm{phot}$. The spectra for the spots and faculae are interpolated with the same surface gravity and metallicity but at different temperatures $T_\mathrm{fac} = T_{\mathrm{eff},*} + \Delta T_\mathrm{fac}$ and $T_\mathrm{spot} = T_{\mathrm{eff},*} - \Delta T_\mathrm{spot}$, where $\Delta T_\mathrm{fac}$ and $\Delta T_\mathrm{spot}$ are used as free parameters in the retrieval.

Additionally, we also considered a cloud layer for our retrieval calculations. Previous studies by \citet{Fisher2018}, \citet{Murgas2020}, \citet{Estrela2021}, and \citet{Ouyang2023} suggested the presence of small aerosols in the atmosphere that cause an upward spectral slope in the optical wavelength range. We therefore included a new power-law cloud model in \textsc{BeAR}, where the optical depth as function of wavelength follows a given power law. The (vertical) wavelength-dependent optical depth of the cloud layer is determined by the exponent of the power law $e_\mathrm{c}$ and the optical depth at a reference wavelength $\tau_\mathrm{c}$:
\begin{equation}
   \tau_\mathrm{c}(\lambda) = \tau_\mathrm{c} \frac{\lambda^{e_\mathrm{c}}}{\lambda_{\mathrm{ref}}^{e_\mathrm{c}}} \ ,
\end{equation}
where $\lambda_{\mathrm{ref}}$ is the reference wavelength for which we used a value of 1\,\textmu m. For a cloud layer that exhibits a Rayleigh scattering-like behaviour, for example, $e_\mathrm{c}$ would be -4. The cloud's location in the atmosphere is determined by a cloud-top pressure $p_\mathrm{c}$ and its vertical extent is assumed to be one atmospheric scale height.

A full list of all priors and their corresponding distributions can be found in Table \ref{tab:retrieval_parameter}. We also tested retrievals with a grey cloud and a clear-sky atmosphere, respectively. The grey cloud model only has two free parameters, the cloud top pressure and the optical depth of the cloud layer.

\begin{table}
  \caption{Summary of retrieval parameters and prior distributions used for the retrieval model.}  
  \label{tab:retrieval_parameter}      
  \centering                                     
  \begin{tabular}{lcc}         
  \hline\hline                       
  Parameter  & \multicolumn{2}{c}{Prior}         \\
             & Type                      & Value \\
  \hline
   $\log g$                        & Gaussian    & $2.726 \pm 0.017$ \\
   $R_\mathrm{p}$                  & uniform     & $0.8 \, R_\mathrm{Jup}$ -- $1.2 \, R_\mathrm{Jup}$ \\
   $R_*$                           & Gaussian    & $0.8129 \pm 0.028 \, R_\odot$ \\
   $x_i$                           & log-uniform & $10^{-12}$ -- $10^{-1}$ \\
   $T$                             & uniform     & 500\,K -- 1800\,K\\
   $\tau_\mathrm{c}$               & log-uniform & $10^{-10}$ -- $10^{2}$ \\
   $e_\mathrm{c}$                  & uniform     & -7 -- 0 \\
   $p_\mathrm{c}$                  & uniform     & $10^{-6}$\,bar -- $5$\,bar\\
   $T_{\mathrm{eff},*}$            & Gaussian    & $4875 \pm 140$ K\\
   $\log g_{*}$                    & Gaussian    & $4.50 \pm 0.15$\\
   $\left[\mathrm{Fe/H}\right]_*$  & Gaussian    & $0.40 \pm 0.04$\\
   $\Delta T_{\mathrm{fac}}$     & uniform     & 0 K -- 500 K\\
   $\Delta T_{\mathrm{spot}}$    & uniform     & 0 K -- 500 K\\
   $f_{\mathrm{fac}}$            & uniform     & 0 -- 1\\
   $f_{\mathrm{spot}}$           & uniform     & 0 -- 1\\
  \hline
  \end{tabular}
\end{table}

\subsection{Simultaneous EulerCam Photometry}
\label{sec:fc-methods}

Using the standard aperture photometry pipeline for EulerCam as described in \cite{Lendl2012} to perform bias, flat and overscan corrections, we obtained EulerCam light curves for each night of FORS2 observations. The aperture photometry is performed for different sets of circular apertures of varying sizes and different reference stars in the field of view. Our final light curves are selected by minimising the photometric scatter between consecutive flux measurements and are shown in Fig. \ref{fig:euler-lcs}. 



To account for any correlated noise arising from instrumental systematics, weather, etc, we fitted the observed light curves with a photometric baseline model along with the pure transit model. The optimal baseline is determined by iteratively fitting different baseline models involving air mass, exposure time, sky background, shifts and FWHM of stellar Point Spread Function (PSF), 
minimising the Bayesian Information Criterion (BIC). For the first night, 2017-07-18, a combination of a second order polynomial of the FWHM of the stellar PSF and first order polynomial of the exposure time is used. For the second night, 2017-08-14, a combination of a first order polynomial of the air mass, a second order polynomial of the FWHM of the stellar PSF and a first order polynomial of the sky background is used. Lastly, for the third night, 2017-08-18, a second order polynomial of the air mass and a first order polynomial of the exposure time are used.

Both the FORS2 and EulerCam light curves taken on the third night show signs of an active region crossing. To model the properties of the hot region, we used an adapted version of the PyTranSpot code, which is designed to model multiband transit light curves showing starspot anomalies \citep{Juvan2018, SAGE24}. We use blackbody radiation to model the emissions from the quiet photosphere and the facula. Since both light curves involve the same facula, we consider the contrast between the facula and the stellar surface in both wavelength bands to be related through a blackbody to the same temperature. This constraint on the contrast ratio between the two bands allows us to constrain the temperature, despite the degeneracy between contrast and surface area that is usually present when fitting active region crossings. We further set the mid-transit time, transit depths, quadratic limb-darkening coefficients, location of active region, size and temperature as jump parameters and provided wide uniform priors for model optimisation. The full set of parameters, their priors and results can be found in Table \ref{tab:facfit}.

\section{Results}
\label{sec:results}
Despite being treated entirely separately, the impact parameter and transit durations obtained from the white light fits are identical during the nights the RI and z grism observations were taken and the B grism impact parameter is consistent within 1\textsigma, while the transit duration is consistent within 2\textsigma. 

\subsection{Transmission spectrum}

The transmission spectra of all three observations are plotted in the top panel of Fig. \ref{fig:spectrum}. An offset is applied between to the B and z grism observations to align them to the RI observations in the overlapping wavelength regions. This is common practice due to differences in systematic instrumental and weather effects between observations, as well as the effect of the common noise model removal. The final values of the planet radius can also be found in Tables \ref{tab:B_slcs}, \ref{tab:RI_slcs}, and \ref{tab:z_slcs}. Errorbars at the blue and red ends of the wavelength range are larger due to the reduced signal-to-noise ratio in the spectra in those regions, which is caused by lower grism throughput. This is especially the case for the bluest wavelength points, where the added lower efficiency of the MIT detector, which was used for all three observations, further increases the errorbars. Because of this, we do not consider the uptick in apparent radius at the three shortest wavelength points significant. 

Generally, the different nights are in good agreement in the overlapping regions. Both the RI and z grism observations show a relatively high transit depth at $\sim$7600\,$\mathrm{\AA}$, but this coincides with the telluric oxygen band and is therefore unlikely to be a planetary feature.

The combined transmission spectrum does not show any obvious atomic or molecular features, indicating a possible cloud cover, but it does show a clear decrease in the apparent planetary radius towards the blue that is characteristic of unocculted faculae.

\subsection{Atmospheric retrievals}

\begin{figure*}
    \centering
    \includegraphics[width=\textwidth]{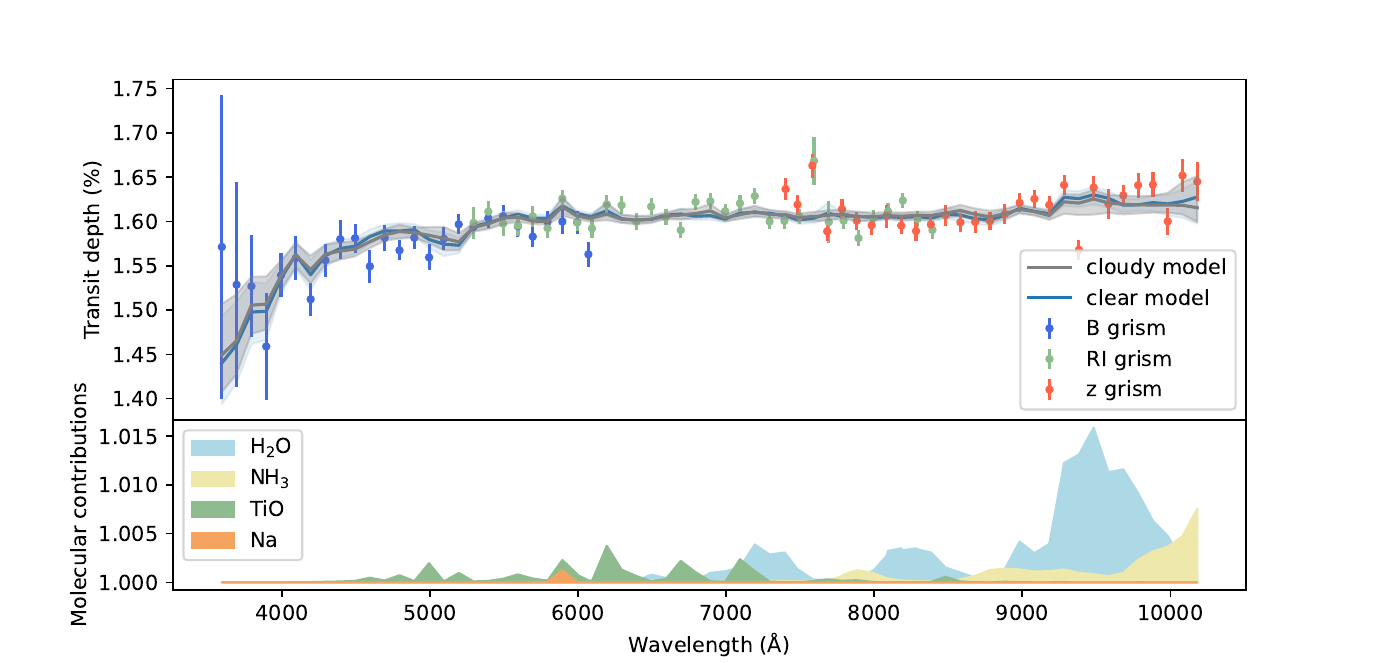}
    \caption{\emph{Top:} the transmission spectrum of WASP-69b. The blue green and red points are the transit depths obtained from the B, RI and z grism observations respectively. The grey line represents the median of all retrieval posterior sample spectra for the favoured power-law cloud model, with the shaded region indicating their 1\textsigma\ interval. The blue line represents the cloud free model. \emph{Bottom:} the relative contributions of the different atoms and molecules to the retrieved spectrum. }
    \label{fig:spectrum}
\end{figure*}
\begin{figure*}
  \includegraphics[width=\textwidth]{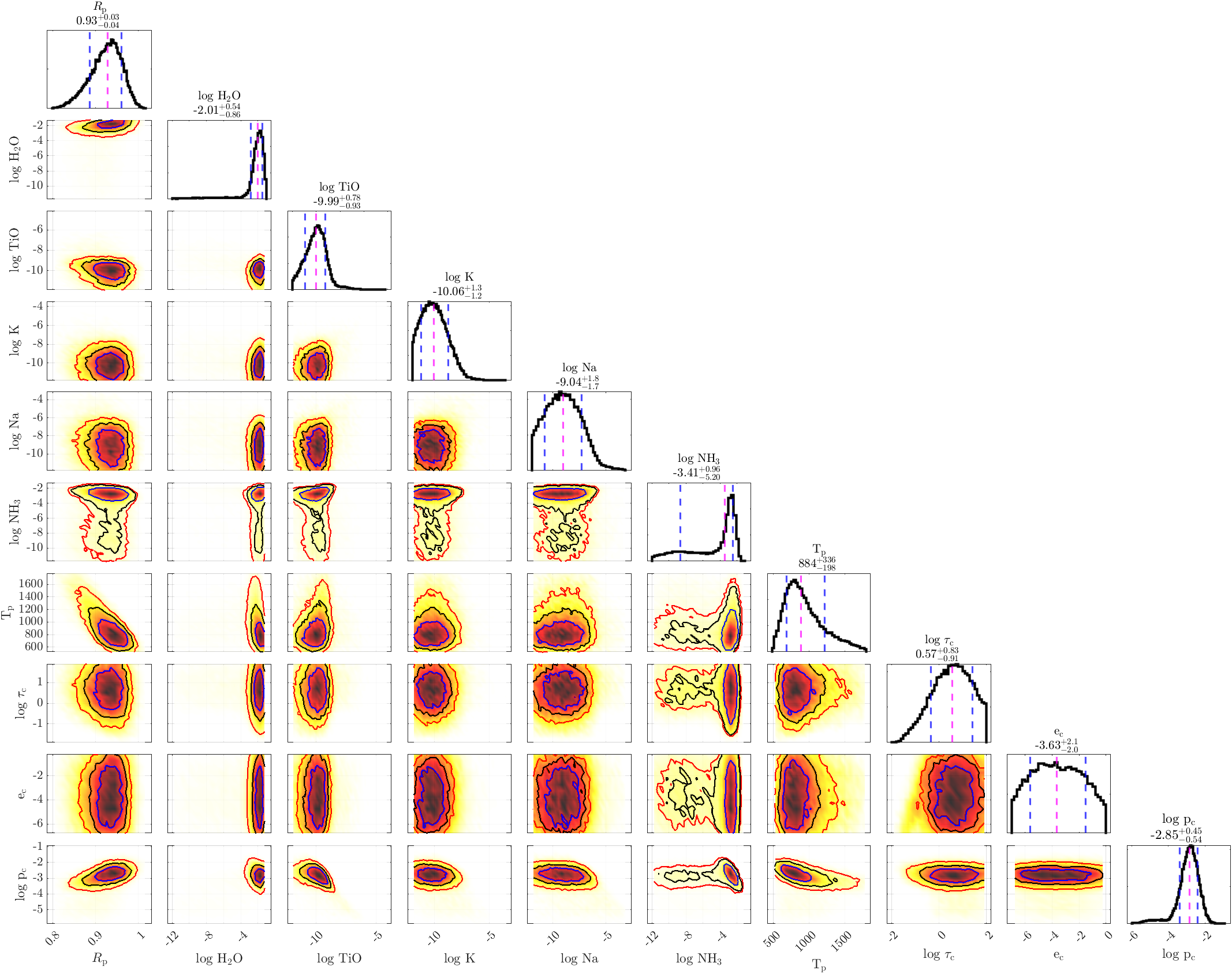} 
  \includegraphics[width=0.5\textwidth]{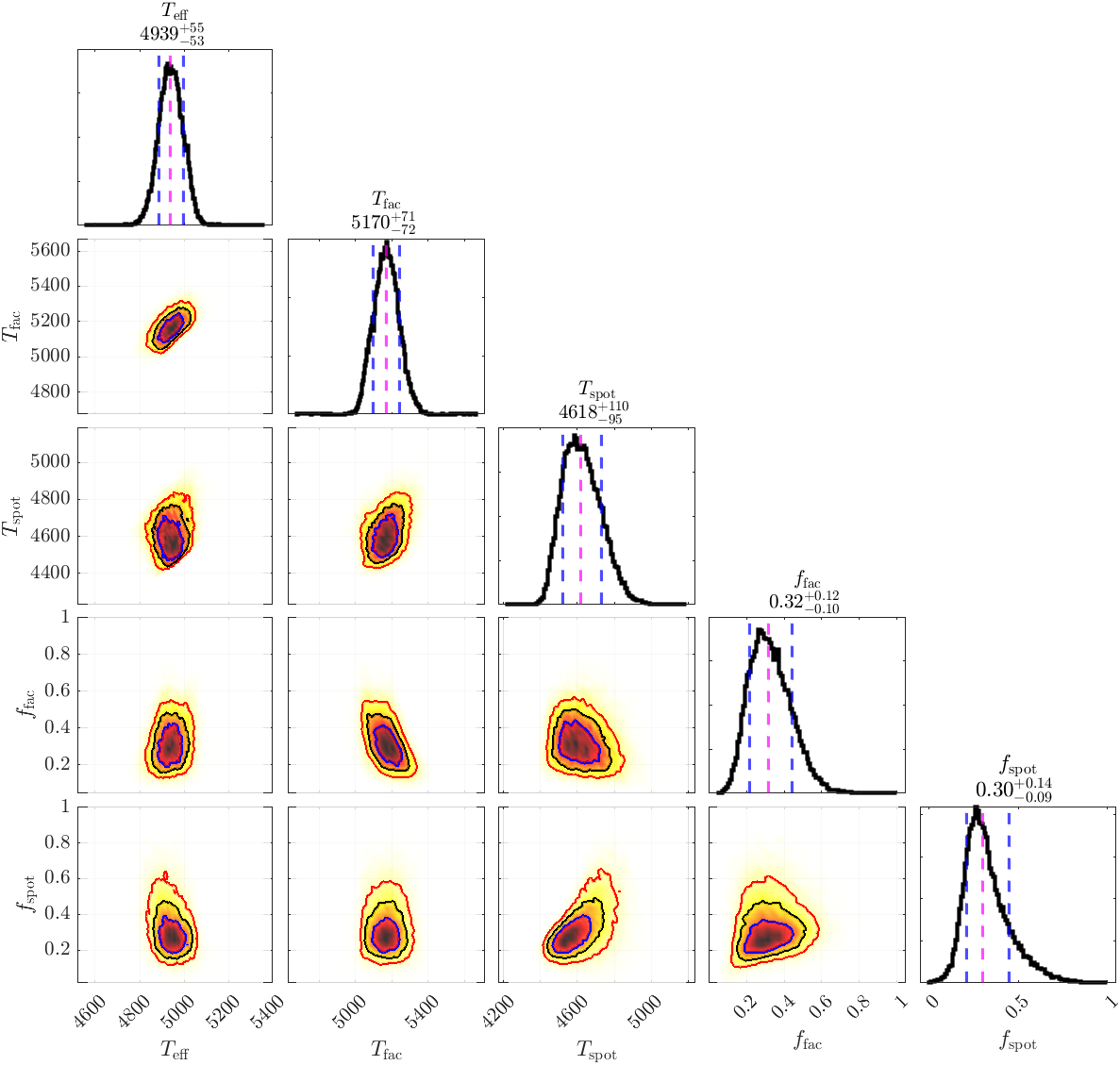} 
  \caption{Posterior distributions for the planet's atmosphere properties (top) and the stellar parameters (bottom). The vertical, magenta lines in the histogram plots refer to the median value of the depicted distribution, while the blue, vertical lines denote the 1\textsigma\ intervals. The three contour lines in the two-dimensional correlation plots refer to the 1\textsigma, 2\textsigma, and 3\textsigma\ regions, respectively.} 
  \label{fig:retrieval_post}
\end{figure*}

The main feature of the transmission spectrum is the drop in apparent radius at the blue end that indicates stellar activity contamination. The median of all posterior spectra for the power-law cloud retrieval is displayed in Fig. \ref{fig:spectrum} in grey, on top of the observational data. The posterior distributions for the stellar activity parameters are shown in the lower panel of Fig. \ref{fig:retrieval_post}. We did not plot the posterior distributions for the star's surface gravity and metallicity since those are essentially sampled from their Gaussian distributions listed in Table \ref{tab:retrieval_parameter}. However, the retrieval prefers a median stellar effective temperature slightly higher than that of its original Gaussian distribution. Our constraints on the temperature of the faculae suggest that they have a median temperature of about 5170$\pm$72\,K, or roughly 230\,K hotter than the fitted stellar effective temperature of 4939$\pm$55\,K. With a median temperature of about 4618$\pm$110\,K, the spots are around 320\,K below the stellar effective temperature. The spot and faculae coverage fractions are both near 30\%.

Beyond the stellar contamination, the spectrum is missing any obvious features, including those of the alkali metals.
As a result, our retrievals suggest the presence of a high-altitude cloud layer in the atmosphere, confirming the findings by, for example, \citet{Fisher2018} or \citet{Estrela2021}. The cloud layer is located at a very low pressure of about 1.4\,mbar with a 1\textsigma\ interval of 0.4-4\,mbar. This very high-altitude cloud is necessary for the retrieval to explain the lack of strong molecular features in the transmission spectrum. In contrast to previous studies we do not find evidence for a Rayleigh-like scattering haze. The power law index $e_\mathrm{c}$ of our cloud model is essentially unconstrained, see upper panel of Fig. \ref{fig:retrieval_post}. This is caused by the strong impact of the stellar contamination at smaller wavelengths, that dominates any upward slope in the transmission spectrum that would be caused by small aerosol particles. We also note that while an upward slope in the transmission spectrum could be caused by small, scattering cloud particles, it could also originate from stellar contamination with predominant spots.

In addition to the power-law cloud, we also performed an additional retrieval test with a grey cloud layer (not shown). This retrieval yields essentially the same posterior distributions for the chemical species and atmospheric temperature, suggesting that a grey cloud is the most probable scenario to describe our observations of WASP-69b. The Bayesian evidence of the grey-cloud model ($\ln \mathcal Z = -613.663$) is only marginally higher than that with the power-law cloud ($\ln \mathcal Z = -614.041$).
The preference for the grey-cloud model is caused by a reduced number of free parameters to explain the data. However, the Bayes factor between the two models is only 1.45 and, thus, statistically insignificant. The relative lack of constraint on the power-law index $e_\mathrm{c}$ and the preference for a grey cloud also means that the cloud particle size cannot be sufficiently constrained.

For comparison, we also performed a cloud-free retrieval. The corresponding posterior distributions can be found in Appendix \ref{sec:appendix_retrieval}. The median model is included in the top panel of Fig. \ref{fig:spectrum} in blue. Without a cloud layer, the abundances of essentially all chemical species shift by at least about an order of magnitude to lower values. Furthermore, the atmospheric temperature becomes prior-dominated at the lower prior boundary of 500 K. Thus, without the cloud layer, the retrieval aims to dampen the molecular features in the transmission spectrum by reducing the temperature, thereby making the atmospheric scale height smaller. The posteriors for the stellar contamination are, on the other hand, roughly similar to the one for the cloudy models. The retrieved temperatures are consistent within their 1\textsigma\ intervals. Only the posterior distribution for $f_\mathrm{fac}$ is slightly more skewed in the cloud-free case, shifting its median value away from the maximum of the distribution. However, both fractional coverages $f_\mathrm{fac}$ and $f_\mathrm{spot}$ are still consistent with those of the cloudy retrieval calculations within their 1\textsigma\ intervals. The cloud-free retrieval model has a substantially lower Bayesian evidence ($\ln \mathcal Z = -619.213$). Compared to the power-law cloud model discussed above, the Bayes factor is about 176, suggesting that a cloudy model is strongly preferred over the clear-sky one.


For the favoured power-law cloud retrieval, our retrieved atmospheric temperature of $T_\mathrm{p}=884^{+340}_{-200}$ K is slightly cooler than the zero-albedo estimate on the equilibrium temperature of $963 \pm18 $ K reported by \citet{Anderson2014}. 

The sodium feature is clearly suppressed due to the aforementioned clouds, which gives us a median mixing ratio of 10$^{-9}$ with a relatively large uncertainty of nearly 2 dex. Additionally, the sodium feature lies in the overlapping region of the B and RI grism wavelengths, meaning there are two data points from different nights at the same wavelength. These points are consistent with each other, but the difference between them still introduces extra uncertainty in the retrieved sodium abundance, which leads to a very weak constraint. This is consistent with previous detections of sodium at high resolution, where the strong linecore can be resolved \citep{Casasayas2017,Deibert2019,Khalafinejad2021}. 
The bottom panel of Fig. \ref{fig:spectrum} shows the contribution of sodium to the final spectrum is small compared to the contributions of the other molecules included in the retrieval. These contributions were obtained by postprocessing the posterior sample and selectively removing a specific absorber. The comparison with the median spectrum of the full posterior sample then yields the relative contribution of the removed species to the full model. 

Similar to sodium, we also do not see a potassium feature. Instead we find a 3\textsigma\ upper limit on the potassium abundance of $<10^{-7}$. Unlike sodium, potassium has not previously been detected in the atmosphere of WASP-69b \citep{Deibert2019,Murgas2020}. It is possible that this is at least in part due to the fact that the potassium feature coincides with a telluric oxygen absorption band. We also put an upper limit on the abundance of titanium oxide of $<10^{-7.7}$ at 3\textsigma\ significance. This means we cannot confirm the hints of TiO seen by \citet{Ouyang2023}, who retrieved abundances of the order of $10^{-2.5}$.

We do constrain the abundance of water and ammonia, with median mixing ratios of about 0.01 and $10^{-3.4}$, respectively. These values are consistent with previous observations of water, including infrared data from HST \citep{Tsiaras2018,Khalafinejad2021,Guilluy2022,Schlawin2024}, and ammonia \citep{Guilluy2022} in the atmosphere of WASP-69b. \citet{Tsiaras2018} do show a posterior for ammonia in their Figure 4 that appears to indicate its presence, but this is not further expanded upon.  
The high abundances of both water and ammonia are driven by the rise in the red end of the transmission spectrum. Both molecules have opacities that increase towards the infrared and therefore have large contributions to the spectrum beyond 7000\,$\mathrm{\AA}$. This can be seen in the bottom panel of Fig. \ref{fig:spectrum}, where the contributions to the transmission spectrum of water are plotted in blue and those of ammonia in yellow. Individual retrievals without water and without ammonia show that the molecules are somewhat degenerate, with neither being preferred over the retrieval with both (Bayesian evidences of 1.8 and 1.14 respectively). However, a retrieval without either molecule is not able to capture the rise and results in a Bayesian evidence of 3.5, equivalent to a detection of 2.2\textsigma. Additionally, while the high abundance of \ch{NH3} especially may seem surprising, recent theoretical, non-equilibrium chemistry calculations for the atmospheric composition of the terminator region of WASP-69b (Bangera et al., in review) suggest that strong vertical mixing of ammonia increases its abundance considerably compared to a chemical-equilibrium case. Their study obtained a mixing ratio of about $10^{-4}$ in the upper part of the atmosphere, which is consistent with our value within its 1\textsigma\ interval. 

\subsection{Active region crossings}
\label{sec:fc-results}
Two of the FORS2 white light curves show potential active region crossings. The B grism light curve shows a bump just before egress that is likely a spot crossing. However, the EulerCam data are of insufficient quality to confirm this, as the correlated noise is of the same amplitude as the potential spot crossing. 

The z grism light curve shows a dip in flux just after ingress. This drop coincides with a change in exposure time due to weather circumstances, although we would expect the use of a reference star to diminish the resulting effects on the light curve. Added credibility is lent to the reality of the feature by the simultaneous EulerCam light curve, which shows a similar drop at the same location during transit. Even more so, the drop is slightly larger, as expected for a facula crossing observed in a bluer bandpass. Since the Euler telescope is located in La Silla and the VLT in Paranal, the atmospheric effects in both light curves are expected to be different and no change in exposure time was required until the egress of the transit, long after the flux drop. Because of this, we consider the drop to be due to the planet crossing a facula on the stellar surface, temporarily increasing the occulted flux. 

\begin{figure*}
      \includegraphics[width=\textwidth]{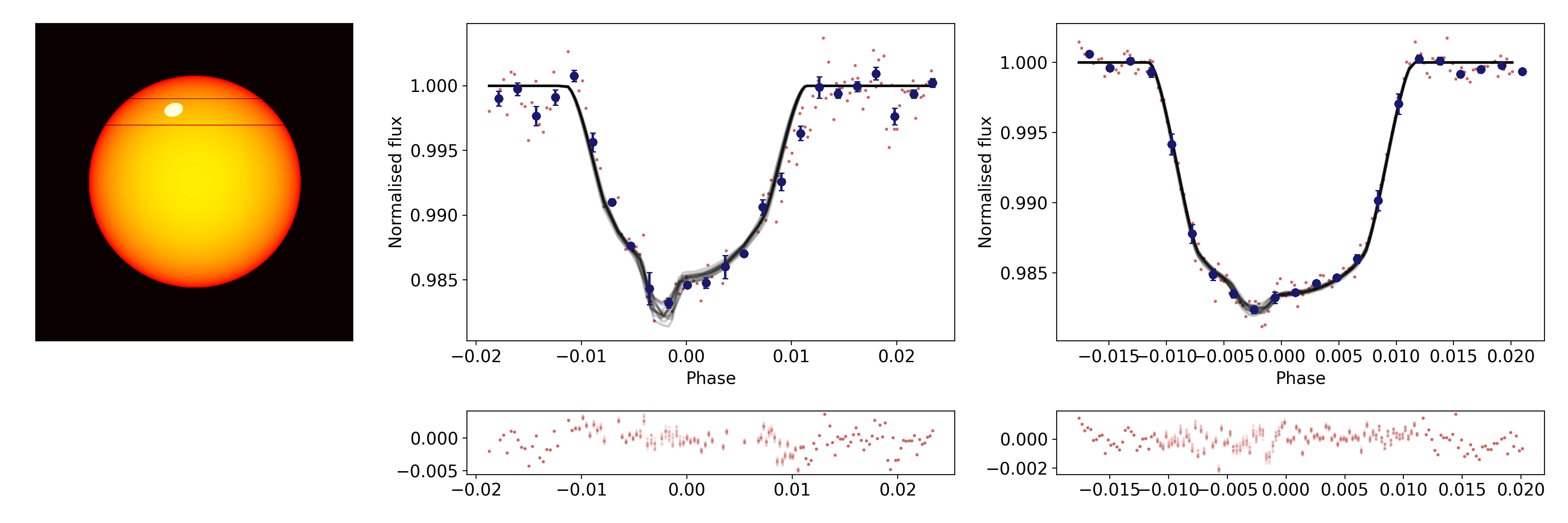} 
      \caption{PyTranSpot models for ECAM and FORS2 observations taken on 2017-08-18. \textit{Left:} A toy model of the stellar surface with the occulted facula and transit chord marked. \textit{Middle and right:} Detrended ECAM and FORS2 white light curves with transit + facula model on top.}
      \label{fig:crossfit}
\end{figure*}

The stellar surface model and the models for both light curves are shown in Fig. \ref{fig:crossfit} and results of the fit in Table \ref{tab:facfit}. 
We find the anomaly in the light curve to be best fit by the planet crossing a facula of 5.1$^{+3.3}_{-2.3}$\degree\ in diameter and with a temperature of $644^{+427}_{-263}$\,K hotter than the rest of the star.



\begin{table}
    \centering
    \begin{tabular}{l l l}
       \hline
       \hline
       Jump parameter & Prior & Results \\
       
       \hline
        Phase offset [10$^{-7}$]  & $\mathcal{U}$(-0.02, 0.02) & 2.67$^{+0.003}_{-0.003}$ \\

        Rp/Rs (EulerCam) & $\mathcal{U}$(0, 0.2) & 0.1177$^{+0.0016}_{-0.0015}$\\

        Rp/Rs (FORS2) & $\mathcal{U}$(0, 0.2) & 0.1246$^{+0.0005}_{-0.0005}$ \\
       
       Latitude [deg]  &  $\mathcal{U}$(0, 90)  & 46.7$^{+6.4}_{-10.5}$  \\
       
       Longitude [deg]  & $\mathcal{U}$(-180, 180) & -15.8$^{+1.5}_{-2.0}$ \\
       
       Size [deg]      & $\mathcal{U}$(1, 10) & 5.1$^{+3.3}_{-2.3}$ \\

       Temperature [Kelvin] & $\mathcal{U}$(4875, 20000) &  5519$^{+427}_{-263}$ \\

       u1 (EulerCam) & $\mathcal{N}$(1.140, 0.1) & 1.09$^{+0.07}_{-0.08}$\\

       u2 (EulerCam) & $\mathcal{N}$(-0.255, 0.1) & -0.33$^{+0.09}_{-0.07}$\\

       u1 (FORS2) & $\mathcal{N}$(0.386, 0.1) & 0.36$^{+0.06}_{-0.06}$\\

       u2 (FORS2) & $\mathcal{N}$(0.209, 0.1) & 0.23$^{+0.08}_{-0.07}$\\

       \hline
    \end{tabular}
    \caption{\label{tab:facfit}Phase offset w.r.t to a mid-transit time of 2457984.62 BJD\_TDB}
\end{table}

\section{Discussion}
\label{sec:disc}

\subsection{Stellar activity changes}
The retrieved activity is dominated by the slope in the B grism part of the spectrum. Since the data were taken over three separate nights, it is possible that the activity has changed between observations. 

The stellar rotation period is 23-24 days \citep{Anderson2014,Khalafinejad2021}. 
Since the B and z grism observations are taken 4 days or \sfrac{1}{6}$^\mathrm{th}$ of a rotation apart and the star is inclined to a nearly pole-on orientation \citep[$\sim65\degree$;][]{SAGE24,Allart2024}, only a small fraction of the surface is expected to have changed and we do not expect significant differences in activity between these observations, either from activity evolution or from the stellar rotation changing what part of the surface is seen. The RI and B grism observations are taken 26 days apart, just over a full rotation. However, cooler stars are expected to have longer lived active regions and Kepler observations of stars with 20-day rotation periods and temperatures have shown active region lifetimes of 20 to 200 days for K stars with similar temperatures to WASP-69 \citep{Giles2017}. While it is possible that the activity has evolved significantly between the RI and B grism observations, it is probably not the case.  

\subsection{Unocculted activity features}

\citet{SAGE24} previously fitted a spot covering fraction of $\sim1$\% for WASP-69 using rotational modulation in the TESS light curve, but this is a lower limit under the assumption that the star is viewed at exactly 90\degree\ inclination. The fact that the system is likely inclined means that there is a large area of the stellar surface that does not rotate out of view. Additional activity can therefore be present without changing the observed intensity variations, which is reflected in the lack of an upper limit on the spot size when fitting the TESS light curve with inclination as a free parameter. In this fit, the inclination comes out to 67$\pm$8\degree.  

There are three previous studies that have obtained medium-resolution transmission spectra of WASP-69 b and analysed possible activity. \citet{Murgas2020} observed WASP-69b with OSIRIS on the GTC in 2016, a year before our own observations. \citet{Estrela2021} combined data from STIS and WFC3 on Hubble, taken in May and October of 2017, two months before and two months after our own observations. Finally, \citet{Ouyang2023} coincidentally observed the same night as our RI grism observations with the 4m SOAR telescope. Their transmission spectrum shows a slope that is not present in ours. Of these three studies, only \citet{Estrela2021} have any data at shorter wavelengths than $\approx500$\,nm, where the effect of unocculted faculae are stronger, and even then only to $\approx$450\,nm. A combination of unocculted faculae and starspots could still introduce a positive slope into the spectrum at their observed wavelengths if the spot coverage fraction were higher than the facula coverage fraction. A difference of 10 percentage-points between the two is enough to introduce a visible slope in the spectrum at the retrieved temperatures, with the slope increasing with the difference. Activity evolution could therefore explain the differences with both \citet{Murgas2020} and \citet{Estrela2021}, despite the latter observations only being a few months removed. Finally, the \citet{Ouyang2023} observations were taken on the same night, but are still limited in the blue wavelength range, making detection of unocculted facula signatures impossible. Additionally, between 560\,nm and 870\,nm our spectrum and that of \citet{Ouyang2023} are generally consistent within 1\textsigma. It is only at the extreme ends of their spectrum that the differences arise, which coincide with an increase structure in the residuals and in the size of the errorbars.  It is possible that differences in the treatment of correlated noise is especially apparent at these wavelengths, as wavelength correlated noise can introduce a slope into the spectrum if it is not fully accounted for \citep{Fortune2024}. 
Additionally, the data from \citet{Estrela2021} was re-analysed by \citet{allen2024}. They found a different slope and concluded they could not distinguish between planet atmosphere and stellar contamination with the precision available in the data. They did not provide any parameters for their stellar contamination model.
An overview of the retrieved possible activity parameters of the first three works is shown in Table \ref{tab:activity}. 

Although all three studies found that their spectra could be fully explained by stellar activity contamination, they ultimately concluded that since no spot crossings have been observed, significant activity is unlikely and have therefore attributed all spectral features to the planet atmosphere. Considering the obvious facular signal in our transmission spectrum, the facula crossing in one of our light curves and the possible spot crossing in another, we cannot conclude the same.

\begin{table*}
    \caption{Retrieved possible activity levels in previous studies compared to this work. Dashes indicate that parameters were not fitted for. Values with no errorbars were held fixed. }
    \renewcommand{\arraystretch}{1.3}
    \label{tab:activity}
    \centering
    \begin{tabular}{lcccccc}
    \hline \hline
    Study & Observation year & $f_{spot}$ & $T_{spot}$(K) & $f_{fac}$ & $T_{fac}$(K) & $T_{phot}$(K) \\
    \hline
    \citet{Murgas2020} & 2016 & 0.55$^{+0.30}_{-0.20}$  & 4594$^{+48}_{-77}$ & 0.15$^{+0.46}_{-0.13}$ & 4788$^{+308}_{-68}$ & 4716$^{+23}_{-33}$ \\
    \citet{Estrela2021} & 2017 & >0.32 & 4604$\pm$24 / 4307$\pm$25 & - & - & 4907$\pm$36 / 4548$\pm$38 \\
    \citet{Ouyang2023} & 2017 & $0.33^{+0.27}_{-0.17}$ & $4305^{+138}_{-297}$ & $0.25^{+0.35}_{-0.17}$ & $4893^{+199}_{-105}$ & 4735 \\
    This work & 2017 & 0.30$^{+0.14}_{-0.09}$ & $4618^{+110}_{-95}$ & $0.32^{+0.12}_{-0.10}$ & 5170$^{+71}_{-72}$ & 4939$^{+55}_{-53}$\\
    \hline
    \end{tabular}
\end{table*}

It is worth noting that the obtained spot and facula coverage fractions and their temperatures are consistent across the four datasets, although this is in large part due to the difficulty of constraining these parameters, which leaves large errorbars. We also find our retrieved parameters for the unocculted faculae to be consistent with the properties obtained from other low-resolution transmission spectra that show facular contamination. These studies find temperatures of 150-350\,K ($\pm\sim$200K) hotter than the stellar surface and coverage fractions of 3-20\% ($\pm\sim$20\%) \citep{Rackham2017,Kirk2021,Jiang2021,Nikolov2021,Cadieux2024}.


\subsection{Crossed facula temperatures}
\label{sec:fc-discussion}
Previous works discussing facula crossings are relatively rare compared to works discussing spot crossings, but include five works discussing transits of single stars and one work that does a statistical analysis on 26 targets \citep{Mohler2013,Kirk2016,Zaleski2019,Zaleski2020,Jiang2021,Baluev2021}. Of the five individually studied stars, three are K-type stars \citep[HATS-2, WASP-52 and HAT-P-12;][]{Mohler2013,Kirk2016,Jiang2021} and the other two are a late G star \citep[Kepler-71;][]{Zaleski2019} and an early M-type star \citep[Kepler-45;][]{Zaleski2020}, both of which are relatively close to K-type stars. \citet{Baluev2021} do not provide spot occurrence rates separated by stellar type. Since WASP-69 is also a K-type star, this suggests there may be an underlying mechanism that makes K-type stars more likely to have bright regions on the surface. 
It is suggested that K-stars are typically spot-dominated \citep[e.g. ][]{Solanki1999,Meunier2024}, but faculae can still exist at the edges of the spots where they can span large areas and last longer than their host spots \citep{Chatzistergos2022}.  

Of the five works discussing individual systems, only \citet{Zaleski2019} provides a temperature for the faculae crossed by their planet, which is 200$\pm$150\,K higher than that of the star for an active region of approximately the same size as the planet, consistent with the 130-350\,K values from unocculted faculae, and similar to the values found for photospheric faculae on the Sun \citep[100 - 500\,K depending on the height in the photosphere; e.g.][]{Frolich2004,Buehler2015,Solovev2019, Pietrow2020,Kuridze2024}. 

With a measured temperature difference of approximately 600\,K, our facula temperature is higher than previous observations, but still consistent with solar values. It is also somewhat higher than the $\sim$200\,K obtained from our measurement of unocculted faculae on WASP-69. However, since that temperature is an average over all the unocculted activity, it is possible that the facula crossing happened to be of a relatively hot region. Additionally, our fit assumes a circular shape for the facula, which is not necessarily accurate given that these features tend to form in between granules \citep[e.g.][]{Joao2023,Danilovic2023}. An elongated shape in the direction perpendicular to the planets path would result in the same crossing time, but a lower contrast area. Increasing the surface area in this way by a factor of two while maintaining the total emission would be possible within the transit chord and would result in temperature difference of approximately 300\,K, consistent with all the above values within 1\textsigma. 


\section{Conclusions}
\label{sec:conc}
We have obtained a ground-based transmission spectrum of WASP-69 from 340 to 1100\,nm by combining three transits observed in different filters with FORS2 on the VLT. 


We observe a clear drop of the transmission spectrum at blue wavelengths, a tell-tale sign of stellar contamination from faculae. We independently confirm this event through the detection of a facula-crossing event observed simultaneously with different facilities. We therefore conclude that even though previous studies dismissed stellar contamination due to a lack of spot crossings, the star is indeed active and this needs to be accounted for. We performed retrievals of combined stellar activity and planet atmosphere models and found spot and facula coverage fractions of approximately 30\%. We also constrained the active region temperatures to $4618\pm110$ for spots and $5170\pm72$ for faculae compared to $4939\pm 55$ for the quiet surface, or -321\,K and +231\,K respectively. The planetary contribution was best described by high-altitude cloud deck, which largely suppressed molecular features. Because of this, we do not constrain the sodium abundance, but we do find indications of the presence of NH$_3$ and H$_2$O and are able to put upper limits on the abundances of TiO and K.

We also obtained simultaneous photometric monitoring data from ECAM on the Euler telescope and detected a facula crossing in both the FORS2 and ECAM data of the third night. This allowed us to break the contrast-area degeneracy and obtain a temperature of $644^{+427}_{-263}$\,K above the stellar effective temperature, although it could be lower for an irregularly-shaped facula rather than a circular one. 

\begin{acknowledgements}

This project has received funding from the European Research Council (ERC) under the European Union’s Horizon 2020 research and innovation programme (project {\sc Four Aces}; grant agreement No 724427). It has also been carried out in the frame of the National Centre for Competence in Research PlanetS supported by the Swiss National Science Foundation (SNSF). D.Eh. and A.De. acknowledge financial support from the SNSF for project 200021\_200726. It has also been carried out in the frame of the National Centre for Competence in Research PlanetS supported by the Swiss National Science Foundation (SNSF). 
AP is supported by the European Research Council (ERC) under the European Union’s Horizon 2020 research and innovationprogramme (grant agreement No. 833251 PROMINENT ERC-ADG 2018).
DE acknowledges financial support from the SNSF for project 200021\_200726. 

This project has been carried out within the framework of the NCCR PlanetS supported by the Swiss National Science Foundation under grants 51NF40\_182901 and 51NF40\_205606. We acknowledge support of the Swiss National Science Foundation under grant number PCEFP2\_194576.

P.E.C. was funded by the Austrian Science Fund (FWF) Erwin Schroedinger Fellowship, program J4595-N.
\end{acknowledgements}

%
%

\bibliographystyle{aa}
\bibliography{main}

\begin{appendix}
    \section{Additional light curves}
    
    \subsection{Euler light curves}

    \begin{figure*}[!h]
        \centering
        \includegraphics[width=\textwidth]{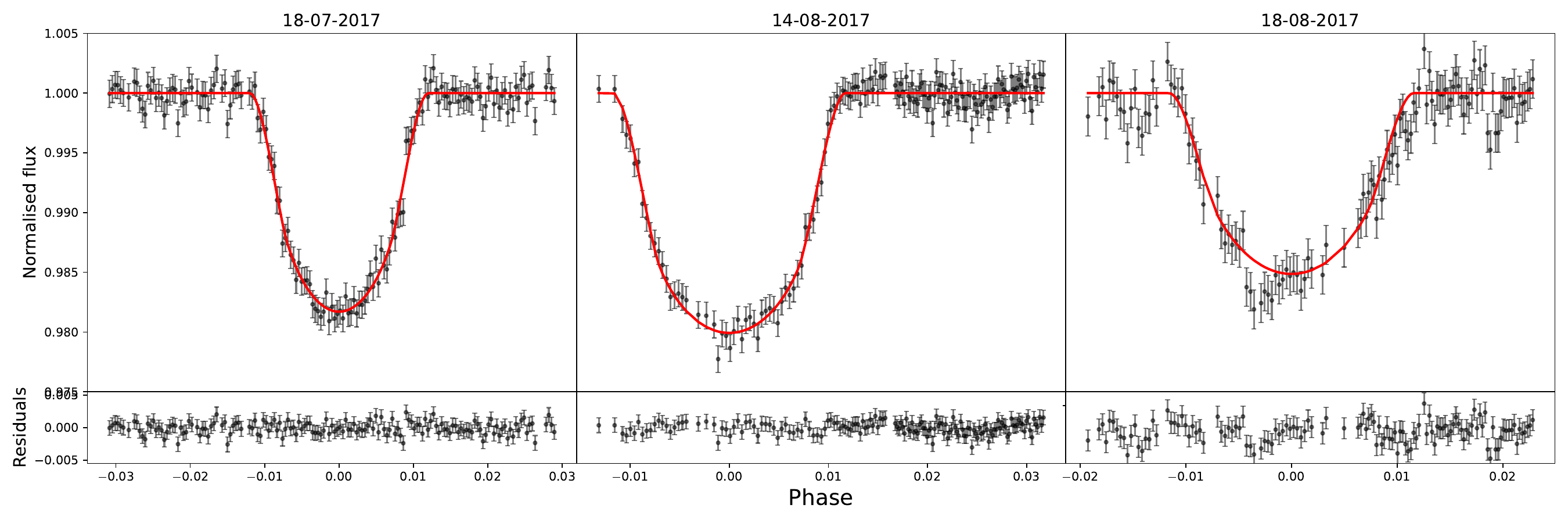}
        \caption{Detrended EulerCam light curves observed simultaneously to the FORS2 data. From left to right, the light curves of the first, second and third night. The top panel show the data after detrending and the transit fit, the bottom panels show the residuals.}
        \label{fig:euler-lcs}
    \end{figure*}

    Figure \ref{fig:euler-lcs} shows the light curves obtained with Euler during the simultaneous observations with FORS2. Due to the smaller size of the telescope, the quality of the data is less than that taken with FORS2, so the light curves of the first and second night are not used in further analysis. 

    All three light curves show significant correlated noise and scatter. The light curve of the first night shows a significant drop in flux just before the transit midpoint, which is due to an increase in atmospheric turbulence. The scatter in the light curve of the second night is too large to confirm or rule out the potential spot crossing observed in the FORS2 data observed on the same night. The light curve of the third night also shows a drop just before the transit midpoint, although a shorter one than the first night and one not caused by weather conditions. Instead, it coincides with the potential facula-crossing in the corresponding FORS2 light curve. Further analysis of this feature in both light curves is provided in Section \ref{sec:fc-methods}. The results and discussion can be found in Sections \ref{sec:fc-results} and \ref{sec:fc-discussion} respectively. 
    

\subsection{Spectroscopic light curves}

\begin{figure*}
    \centering
    \includegraphics[width=\textwidth]{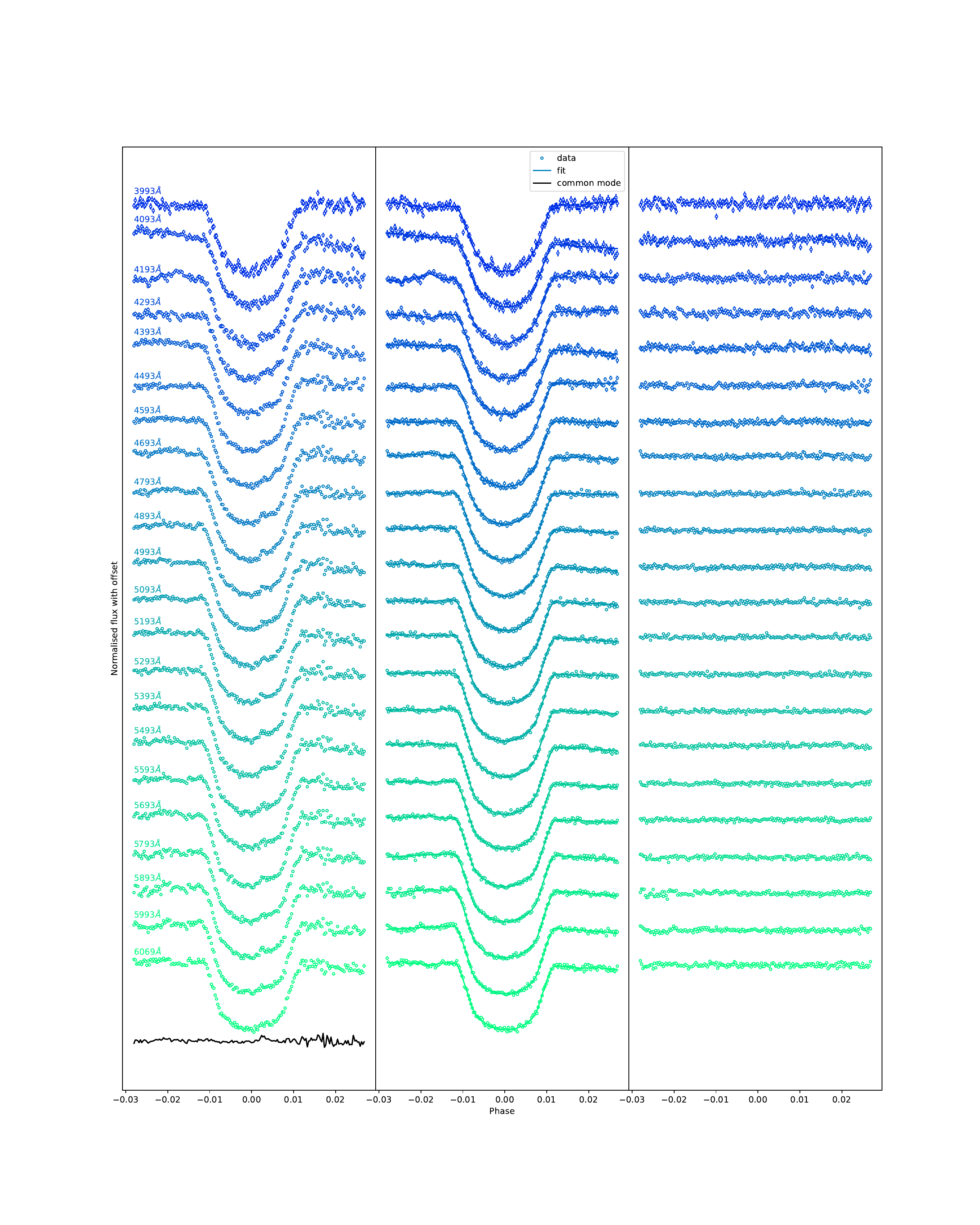}
    \caption{Spectroscopic light curves observed on 15-08-2017 in the B grism. The raw light curves are plotted in the left panel in color, with the shortest wavelengths at the top (bluest) and the longest at the bottom (greenest). The common mode as determined from the white light curve is plotted at the bottom in black. The light curves with the common mode removed are plotted in the middle panel with the fitted transits overplotted in the same color. The residuals are shown in the right panel.}
    \label{fig:spectroscopic-B}
\end{figure*}

\begin{figure*}
    \centering
    \includegraphics[width=\textwidth]{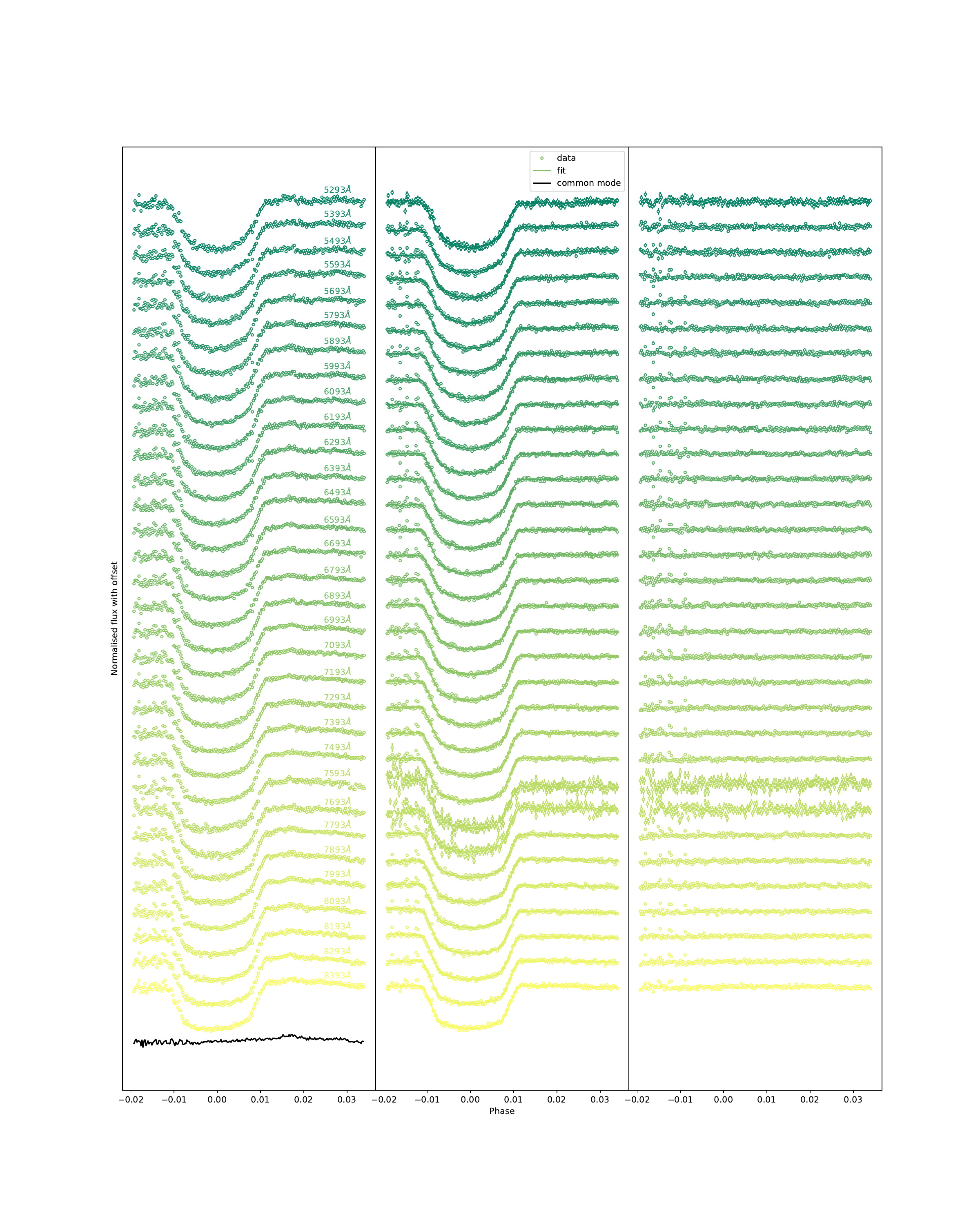}
    \caption{Spectroscopic light curves observed on 19-07-2017 in the RI grism. The raw light curves are plotted in the left panel in color, with the shortest wavelengths at the top (greenest) and the longest wavelengths at the bottom (yellowest). The common mode as determined from the white light curve is plotted at the bottom in black. The light curves with the common mode removed are plotted in the middle panel with the fitted transits overplotted in the same color. The residuals are shown in the right panel.}
    \label{fig:spectroscopic-RI}
\end{figure*}

\begin{figure*}
    \centering
    \includegraphics[width=\textwidth]{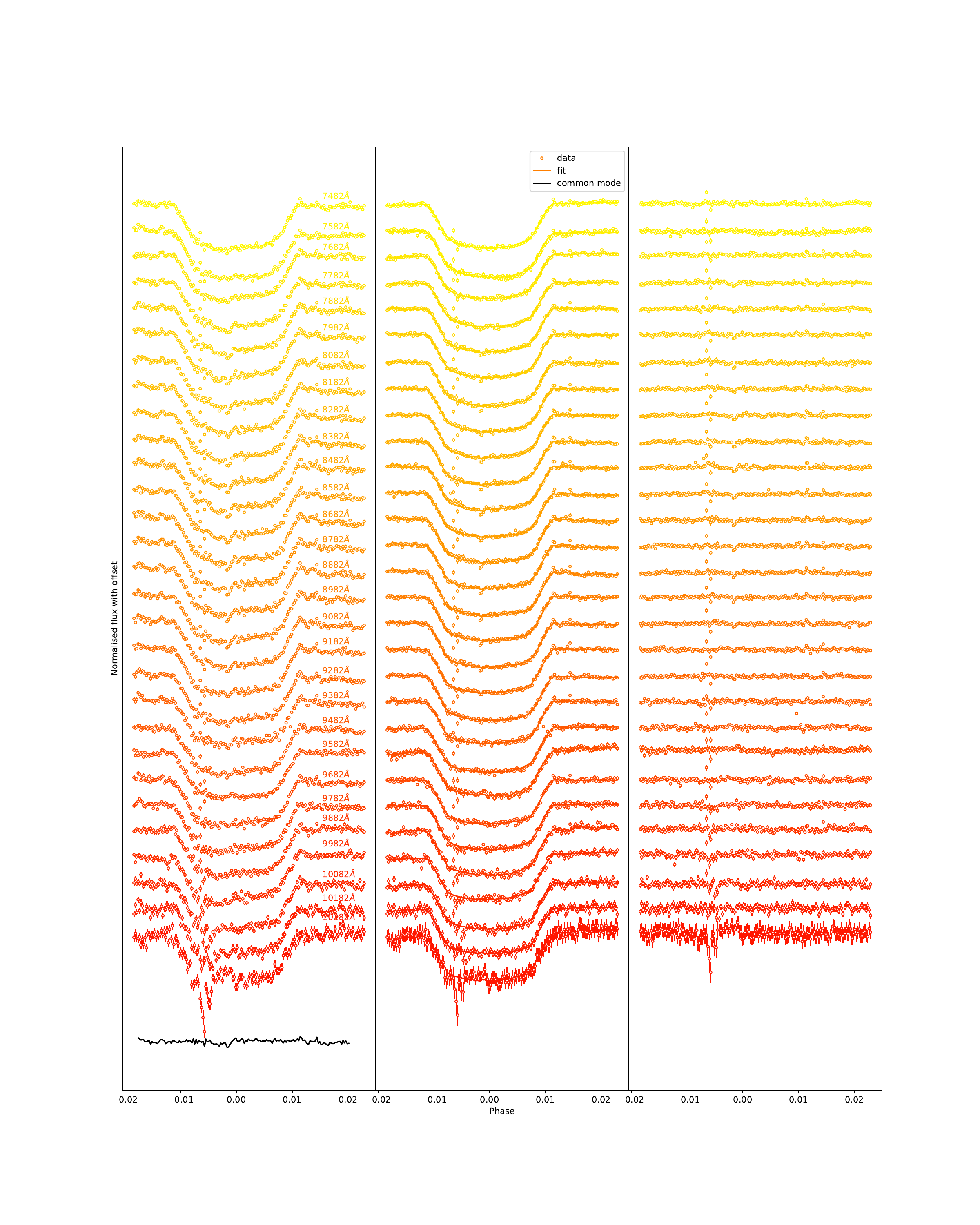}
    \caption{Spectroscopic light curves observed on 19-08-2017 in the z grism. The raw light curves are plotted in the left panel in color, with the shortest wavelengths at the top (yellowest) and the longest wavelengths at the bottom (reddest). The common mode as determined from the white light curve is plotted at the bottom in black. The light curves with the common mode removed are plotted in the middle panel with the fitted transits overplotted in the same color. The residuals are shown in the right panel.}
    \label{fig:spectroscopic-z}
\end{figure*}

Figures \ref{fig:spectroscopic-B}, \ref{fig:spectroscopic-RI} and \ref{fig:spectroscopic-z} show the spectroscopic binned light curves for the B, RI and z grisms. All light curves are binned to 10\,nm along the wavelength axis. The left columns show the raw, uncorrected light curves, aranged from shortest wavelengths at the top to longest at the bottom of the plot, with the common noise model plotted in black at the very bottom. The middle columns show the corrected light curves with the fitted transit and baseline model overplotted in the same colour. Finally, the rightmost columns show the residuals for each light curve. 

While some structure remains in some of the residuals these are accounted for in the red and white noise factors used to inflate the errorbars before the final fit of the light curves. The scatter in the light curves is the largest for the bluest B grism light curves and the reddest z grism light curves, where the spectral intensity is lowest. Additionally, there are two light curves at 7693\,$\AA$ and 7893\,$\AA$ in the RI grism where the scatter increases after the correction. These light curves correspond to the telluric oxygen feature. 

The fitted transit depths for the light curves of all three nights are given in Tables \ref{tab:B_slcs}, \ref{tab:RI_slcs} and \ref{tab:z_slcs}. 

\begin{table}
\caption{Spectroscopic fit results for the B grism. The first column contains the wavelength of the binned light curve, the second the fitted planetary radii in units of the stellar radius. }
\label{tab:B_slcs}
\centering
\begin{tabular}{ll}
\hline \hline 
$\lambda$[nm] & $R_p/R_*$ \\
\hline
3493 & 0.08219$\pm$0.03077 \\
3593 & 0.13111$\pm$0.00638 \\
3693 & 0.12947$\pm$0.00441 \\
3793 & 0.12941$\pm$0.0022  \\
3893 & 0.12675$\pm$0.00239 \\
3993 & 0.12988$\pm$0.00095 \\
4093 & 0.13063$\pm$0.00094 \\
4193 & 0.12883$\pm$0.00074 \\
4293 & 0.13053$\pm$0.00072 \\
4393 & 0.13144$\pm$0.00082 \\
4493 & 0.13148$\pm$0.00062 \\
4593 & 0.13027$\pm$0.00071 \\
4693 & 0.1315$\pm$0.00056  \\
4793 & 0.13096$\pm$0.00044 \\
4893 & 0.1315$\pm$0.0005   \\
4993 & 0.13066$\pm$0.00056 \\
5093 & 0.13146$\pm$0.0005  \\
5193 & 0.13208$\pm$0.00043 \\
5293 & 0.13196$\pm$0.0004  \\
5393 & 0.13234$\pm$0.00041 \\
5493 & 0.13242$\pm$0.0005  \\
5593 & 0.132$\pm$0.00048   \\
5693 & 0.13155$\pm$0.00045 \\
5793 & 0.13219$\pm$0.00046 \\
5893 & 0.13219$\pm$0.00052 \\
5993 & 0.13217$\pm$0.00049 \\
6069 & 0.13079$\pm$0.00052 \\
\hline
\end{tabular}
\end{table}

\begin{table}
\caption{Spectroscopic fit results for the RI grism. The first column contains the wavelength of the binned light curve, the second the fitted planetary radii in units of the stellar radius. }
\label{tab:RI_slcs}
\centering
\begin{tabular}{ll}
\hline \hline 
$\lambda$[nm] & $R_p/R_*$ \\
\hline
5193 & 0.1248$\pm$0.00084  \\
5293 & 0.12645$\pm$0.00071 \\
5393 & 0.12693$\pm$0.00047 \\
5493 & 0.12639$\pm$0.00057 \\
5593 & 0.12632$\pm$0.00046 \\
5693 & 0.12671$\pm$0.00046 \\
5793 & 0.12618$\pm$0.00045 \\
5893 & 0.12751$\pm$0.00041 \\
5993 & 0.12644$\pm$0.0004  \\
6093 & 0.12616$\pm$0.00041 \\
6193 & 0.12725$\pm$0.0004  \\
6293 & 0.12721$\pm$0.00041 \\
6393 & 0.12649$\pm$0.00037 \\
6493 & 0.12717$\pm$0.0004  \\
6593 & 0.12668$\pm$0.00036 \\
6693 & 0.12609$\pm$0.00035 \\
6793 & 0.12737$\pm$0.00036 \\
6893 & 0.12739$\pm$0.00036 \\
6993 & 0.12695$\pm$0.00032 \\
7093 & 0.12728$\pm$0.00035 \\
7193 & 0.12762$\pm$0.00034 \\
7293 & 0.12649$\pm$0.00036 \\
7393 & 0.12652$\pm$0.00036 \\
7493 & 0.12673$\pm$0.00034 \\
7593 & 0.12919$\pm$0.00105 \\
7693 & 0.12646$\pm$0.00094 \\
7793 & 0.12654$\pm$0.00037 \\
7893 & 0.12576$\pm$0.00036 \\
7993 & 0.12665$\pm$0.00038 \\
8093 & 0.12696$\pm$0.00036 \\
8193 & 0.12742$\pm$0.00033 \\
8293 & 0.1266$\pm$0.00036  \\
8393 & 0.1261$\pm$0.00042  \\
8447 & 0.12013$\pm$0.00236 \\
\hline
\end{tabular}
\end{table}

\begin{table}
\caption{Spectroscopic fit results for the z grism. The first column contains the wavelength of the binned light curve, the second the fitted planetary radii in units of the stellar radius. }
\label{tab:z_slcs}
\centering
\begin{tabular}{ll}
\hline \hline 
$\lambda$[nm] & $R_p/R_*$ \\
\hline
7371  & 0.1272$\pm$0.00048  \\
7482  & 0.1265$\pm$0.00044  \\
7582  & 0.12824$\pm$0.00053 \\
7682  & 0.12529$\pm$0.00043 \\
7782  & 0.1263$\pm$0.00044  \\
7882  & 0.12577$\pm$0.0004  \\
7982  & 0.12557$\pm$0.00041 \\
8082  & 0.12598$\pm$0.00051 \\
8182  & 0.12555$\pm$0.0004  \\
8282  & 0.12531$\pm$0.00044 \\
8382  & 0.1256$\pm$0.0004   \\
8482  & 0.12603$\pm$0.00047 \\
8582  & 0.12571$\pm$0.00043 \\
8682  & 0.12572$\pm$0.00049 \\
8782  & 0.12575$\pm$0.0004  \\
8882  & 0.12608$\pm$0.00044 \\
8982  & 0.12658$\pm$0.00043 \\
9082  & 0.12676$\pm$0.0004  \\
9182  & 0.12648$\pm$0.00041 \\
9282  & 0.12738$\pm$0.00043 \\
9382  & 0.12446$\pm$0.00046 \\
9482  & 0.12727$\pm$0.00049 \\
9582  & 0.12653$\pm$0.00067 \\
9682  & 0.12691$\pm$0.00049 \\
9782  & 0.12737$\pm$0.00056 \\
9882  & 0.12739$\pm$0.00058 \\
9982  & 0.12575$\pm$0.00061 \\
10082 & 0.1278$\pm$0.00072  \\
10182 & 0.12752$\pm$0.00087 \\
10282 & 0.1288$\pm$0.00188  \\
10382 & 0.12265$\pm$0.00422 \\
10482 & 0.1193$\pm$0.00372  \\
10552 & 0.12925$\pm$0.00873 \\
\hline
\end{tabular}
\end{table}

\section{Cloud-free atmospheric retrieval}
\label{sec:appendix_retrieval}

\begin{figure*}
  \includegraphics[width=0.7\textwidth]{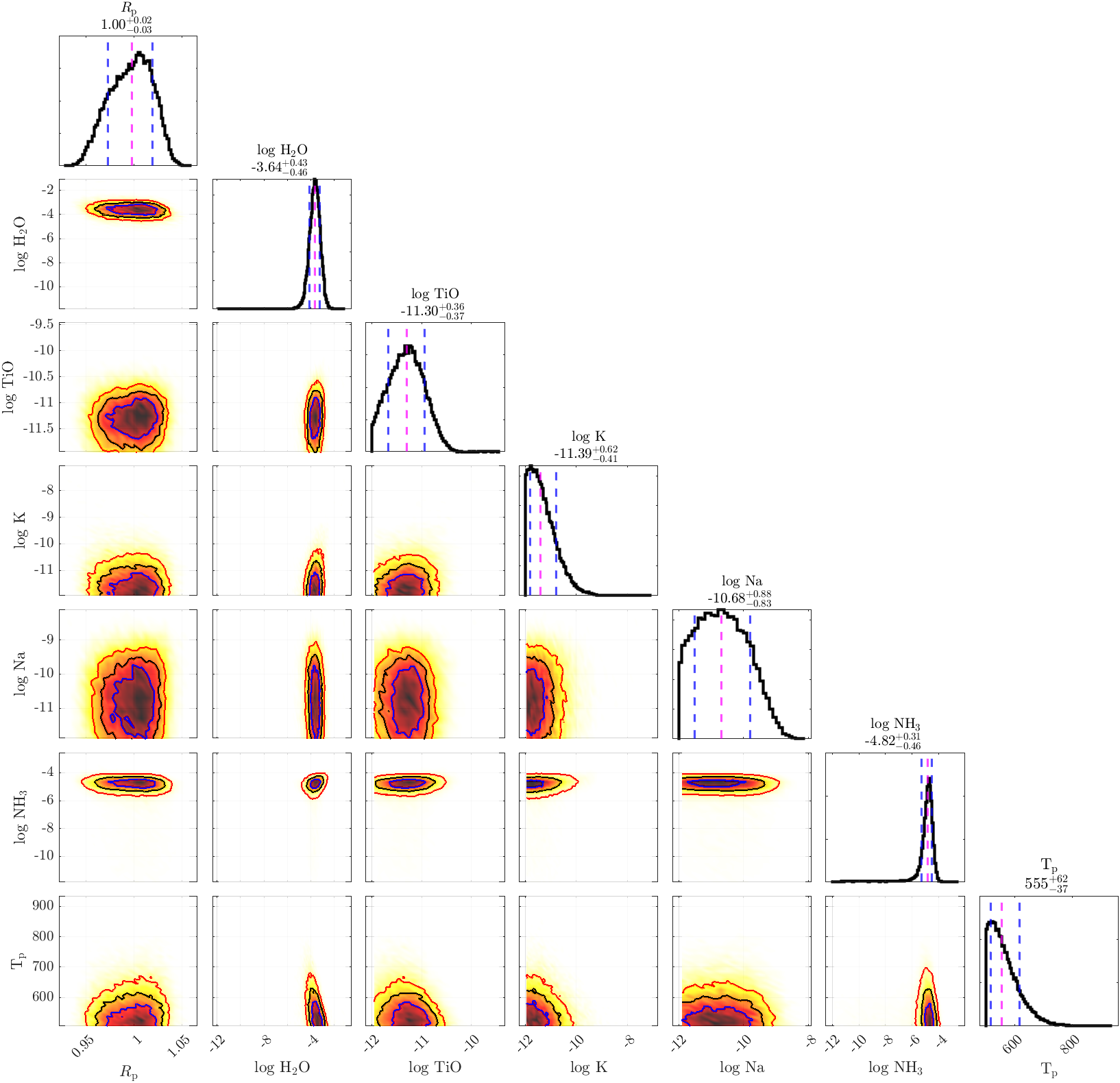}\\ 
  \includegraphics[width=0.5\textwidth]{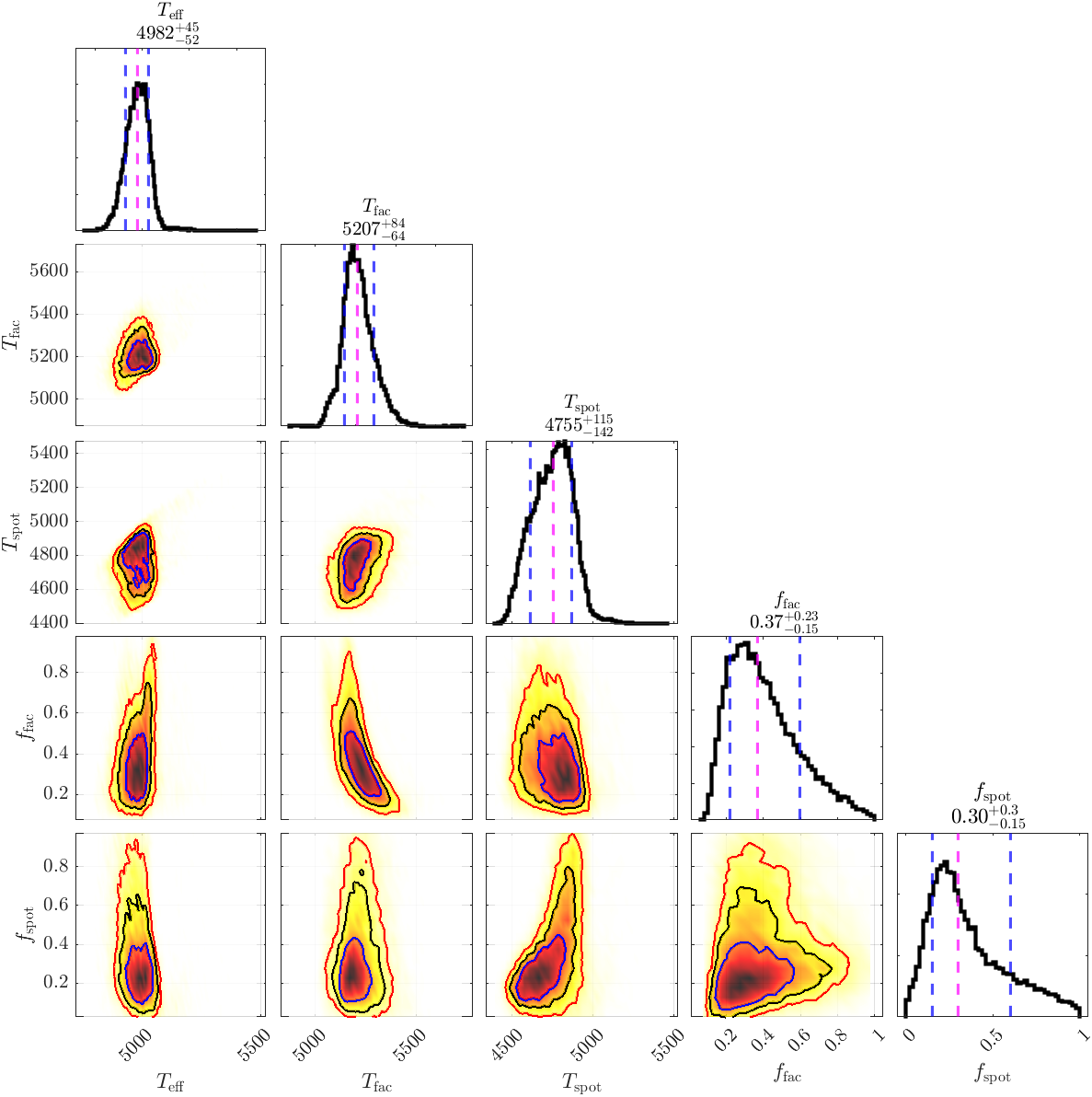} 
  \caption{Posterior distributions for the planet's atmosphere properties (top) and the stellar parameters (bottom) for the cloud-free retrieval model. The vertical, magenta lines in the histogram plots refer to the median value of the depicted distribution, while the blue, vertical lines denote the 1\textsigma\ intervals. The three contour lines in the two-dimensional correlation plots refer to the 1\textsigma, 2\textsigma, and 3\textsigma\ regions, respectively.} 
  \label{fig:retrieval_post_clearsky}
\end{figure*}

A retrieval with a cloud-free planetary atmosphere was run for comparison with the cloudy scenario. The posteriors for this scenario are shonw in Fig. \ref{fig:retrieval_post_clearsky}. Similar to the cloudy case, the posteriors for the surface gravity and metallicity are not shown, since they are essentially sampled from the Gaussian distributions. The resulting spectrum is shown in Fig. \ref{fig:spectrum}. 

\end{appendix}

\end{document}